\documentclass{JHEP3}
\usepackage[xdvi]{graphicx} 
\usepackage{amsmath}
\usepackage{enumerate}
 
\title{A new recipe for causal completions}
\author{ Donald Marolf\footnote{E-mail: {\tt marolf@physics.syr.edu}}\ and
Simon F. Ross\footnote{E-mail: {\tt S.F.Ross@durham.ac.uk}}\\
$^*$ Physics Department, Syracuse University, Syracuse, New York
13244 USA\\
$^\dagger$ Centre for Particle Theory, Department of Mathematical
Sciences, University of Durham, South Road, Durham DH1 3LE UK}

\date{February 2003}

\abstract{ We discuss the asymptotic structure of spacetimes,
presenting a new construction of ideal points at infinity and
introducing useful topologies on the completed space.  Our
construction is based on structures introduced by Geroch, Kronheimer,
and Penrose and has much in common with the modifications introduced
by Budic and Sachs as well as those introduced by Szabados.  However,
these earlier constructions defined ideal points as equivalence
classes of certain past and future sets, effectively defining the
completed space as a quotient.  Our approach is fundamentally
different as it identifies ideal points directly as appropriate pairs
consisting of a (perhaps empty) future set and a (perhaps empty) past
set.  These future and past sets are just the future and past of the
ideal point within the original spacetime.  This provides our
construction with useful causal properties and leads to more
satisfactory results in a number of examples.  We are also able to
endow the completion with a topology.  In fact, we introduce two
topologies, which illustrate different features of the causal
approach. In both topologies, the completion is the original
spacetime together with endpoints for all timelike curves.  We
explore this procedure in several examples, and in particular for
plane wave solutions, with satisfactory results.  }  \keywords{Causal
structure, conformal boundary, plane waves} \preprint{SUGP-03/3-1,
DCTP-03/09}

\begin{document}

\newtheorem{thm}{Theorem}
\newtheorem{defn}{Definition}
\newtheorem{lemma}{Lemma}
\newtheorem{corr}{Corollary}

\section{Introduction}

In many situations, important physical questions about spacetimes are
concerned with what happens `at the boundary.' For example, in
cosmology the classical spacetime modeling our universe has an
initial singularity, and it is important to understand how a more
complete theory assigns initial conditions in the region where
curvatures become large. In scattering experiments, the
experimentalist specifies a collection of ingoing particles at large
distances and early times, and measures a collection of outgoing
particles at late times. More generally, to study non-compact
spacetimes in quantum theories of gravity, we must impose some
restrictions on the asymptotic behaviour of the metric and other
fields. In all these cases, the spacetime is a manifold, and so does
not have a real boundary. A technique which allowed us to equip a
spacetime $M$ with a boundary in some abstract sense, defining some
suitable completion $\bar{M}$, would provide a useful tool in
addressing such problems.

This is a long-studied issue; Penrose provided the first successful
construction of such a boundary in his characterisation of
asymptotically simple spacetimes~\cite{Penrose} almost 40 years
ago. Asymptotically simple spacetimes model the behaviour of isolated
weakly-gravitating sources. A manifold $M$ with metric $g$ is said to
be asymptotically simple if there exists a strongly causal space
$(\tilde{M}, \tilde{g})$ and an imbedding $\theta: M \to \tilde{M}$
which imbeds $M$ as a manifold with smooth boundary $\partial M$ in
$\tilde{M}$, such that:
\begin{enumerate}
\item There is a smooth function $\Omega$ on $\tilde{M}$ such that
  $\Omega >0$ and $\Omega^2 g = \theta_*(\tilde{g})$ on $\theta(M)$,
\item $\Omega =0$ and $d \Omega \neq 0$ on $\partial M$,
\item every null geodesic in $M$ has two endpoints on $\partial M$. 
\end{enumerate}
This definition associates a boundary with $M$ in a well-defined
mathematical sense. Furthermore, it extends the conformal structure on
$M$ (the metric up to conformal transformations) to a conformal
structure on $\bar{M} = M + \partial M$. Thus, we can say quite a bit
about how a particular point on $\partial M$ is related to points of
$M$; it is these relations which are important for studying physical
questions.

In~\cite{Geroch}, Geroch, Kronheimer and Penrose (GKP) proposed a
different approach to adding a boundary to spacetime. This approach,
which will be reviewed in more detail in section~\ref{rev}, is based
on the causal structure of spacetime. Basically, one characterises
points of the spacetime in terms of the timelike curves which end on
them. One can then extend the spacetime by adding endpoints to
timelike curves for which they do not already exist.  In particular,
two curves $\gamma_1$, $\gamma_2$ are given the same future endpoint
if $I^-[\gamma_1] = I^-[\gamma_2]$ (and similarly with past and future
interchanged). This approach is based on the causal structure of
spacetime, so we only get causal (and not conformal) relations between
the new boundary points, dubbed `ideal points', and the original
spacetime. This relaxing of the structure gets us several advantages
over asymptotic simplicity in exchange. The causal approach yields a
procedure for constructing the boundary, whereas the conformal
approach provides no algorithm for constructing an appropriate
$(\tilde{M}, \tilde{g})$ satisfying properties 1-3 above. In addition,
the causal approach may be applied to arbitrary strongly causal
spacetimes, whereas the conformal approach requires conformal flatness
at large distances -- a condition which covers many spacetimes of
physical interest but remains a severe restriction on the space of all
metrics.

However, there is an important technical difficulty with the causal
approach.  In~\cite{Geroch}, it was pointed out that we cannot regard
all the future endpoints we add to curves as distinct from all the
past endpoints; some ideal points can act as both the future and past
endpoints of timelike curves. Thus, GKP~\cite{Geroch} and succeeding
authors suggested that identifications be made between the future and
past endpoints, and that the ideal points should be equivalence
classes generated by some relation. 

GKP~\cite{Geroch} proposed that these identifications should be
obtained by first introducing a topology and then imposing the minimum
set of identifications necessary to obtain a Hausdorff space. However,
this construction fails to produce the `obvious' completion in some
examples~\cite{diff1,diff2}. Attempts have been made in the past to
modify the topology to produce more satisfactory results~\cite{racz},
but these modifications also fail in some examples~\cite{diff3}.  In
part, this is because the introduction of an appropriate topology is
itself problematic in the causal approach.  An approach to the
identifications based more directly on the causal structure was
introduced by Budic and Sachs~\cite{budic} and further refined by
Szabados~\cite{szab1,szab2}.  We review Szabados' rule for performing
identifications in section~\ref{iden}, using a slightly different
language and notation.\footnote{An interesting and different approach
based on the causal structure was recently adopted
in~\cite{garcia}. Their approach is more axiomatic, and will therefore
suffer from some of the same practical drawbacks as Penrose's asymptotic simplicity, though it in principle
applies to a much larger class of spacetimes. With the causality
introduced in \cite{szab1} ($\prec_C$ below), the GKP construction
with Szabados identifications yields a particular procedure for
implementing the axiomatic approach of \cite{garcia} which produces a
unique result.} These causal approaches produce more `natural'
identifications, reproducing the intuitive point-set structure for
$\bar M$ in various examples.

However, the construction of $\bar M$ from equivalence classes suffers
from a fundamental difficulty: it has not been possible to give a
general definition of the timelike pasts and futures of points in
$\bar M$ which will always agree with that on $M$.\footnote{Szabados
claimed in~\cite{szab1} that such a definition was possible for his
$\bar M$, but we exhibit an example where this definition fails in
appendix \ref{app:ex}.} The identifications can introduce causal
relations between points of $M$ not present in the original spacetime,
thus altering the structure on which the causal approach is itself
based.  The problem arises when one attempts to turn a natural
relation $R_{pf}$ between past and future endpoints of curves into an
equivalence relation by enforcing transitivity by hand.

These approaches thus fail in an essential way. If we compare the
logic of the causal approach to that of the conformal approach,
requiring that the notions of $I^\pm(p)$ extend from $M$ to $\bar M$
is analogous to the requirement that $\tilde M$ have a metric
conformal to the metric on $M$. Thus, we feel that any construction
which fails to satisfy this property should not properly be described
as a causal completion. This suggests that radical revision of the
construction of $\bar M$ may be necessary. 

In section~\ref{pairs}, we will propose a new approach to the
construction of $\bar M$, which is not based on a quotient. In our
approach, points of $\bar M$ are constructed from naturally related
IP-IF pairs. The idea is that every point of $M$ has both a past and a
future, and while considering just one or the other uniquely specifies
the point, it does not fully exploit the information in the causal
structure. Thus, we adopt an approach where points of $\bar M$ are
specified by giving both their past and future (for ideal points, one
or the other may be empty). We construct pairs from IPs and IFs that
are related through the same relation $R_{pf}$ used
in~\cite{szab1}. Thus, the key difference between our approach and
previous approaches is that we avoid completing this to an equivalence
relation---a step that has nothing to do with the causal structure.

The IP-IF pair directly determines the past and future of the
associated point of $\bar M$. This tighter connection between the
ideal points and the primitive structures allows us to prove that the
completion $\bar M$ will always have a chronology relation (i.e., a
notion of timelike past and futures) which is compatible with that on
$M$. Thus, this definition of $\bar M$ satisfies what we regard as the
key criterion for a causal completion. The necessary chronology
relation is discussed in section~\ref{chron}. We also discuss the
construction of a causality on $\bar{M}$ in section \ref{caus}.  Here
the situation is somewhat less satisfactory, as different apparently
reasonable notions of causality present themselves. However, as the
construction is based on timelike, and not causal, pasts and futures,
we view the causality as less important than the chronology. 

Another important issue is the construction of a suitable topology on
$\bar M$. Since we want to use the ideal points to represent
asymptotic, or limiting, behaviour at large distances in spacetime, we
need a suitable topology to tell us what limits in the bulk spacetime
$M$ approach which ideal points in $\bar M$. Of course, we want these
limits to be compatible with our causal structure. We therefore want
to introduce a topology on $\bar{M}$ such that $\bar{M}$ corresponds
to $M$ equipped with a suitable boundary in the topological sense, and
such that the TIP $I^-[\gamma]$ and TIF $I^+[\gamma]$ are limit points
of the curve $\gamma$. While the original approach of Geroch,
Kronheimer and Penrose~\cite{Geroch} was based on the further
assumption that the topology was Hausdorff, this strong assumption is
sacrificed in the approaches based more directly on
causality~\cite{budic,szab1,szab2}.

In the quotient constructions, the topology adopted was essentially
that introduced by Geroch, Kronheimer and Penrose~\cite{Geroch},
perhaps with small refinements\footnote{An exception is that of \cite{budic}, which is similar
in spirit to the one we introduce, but which is limited to causally continuous spacetimes.}. This topology is basically constructed
by introducing a suitable `generalised Alexandrov' topology on the
space of all IPs and IFs, and adopting the quotient topology on the
space $\bar{M}$ obtained after identifying the members of an
equivalence class. This construction allows one to define a topology
satisfying the basic requirements above, but it does not always give
the expected relations between limits and ideal points in simple
examples. Some examples of these difficulties were discussed
in~\cite{diff1,diff2,diff3}, and similar issues arise even in de Sitter
space and anti-de Sitter space.

Our change from identifications to pairs requires a new definition of
the topology. The construction of a topology remains a difficult and
subtle issue. Given our focus on chronology, one might think the most
natural thing to do is to adopt an Alexandrov topology, based on the
timelike pasts and futures defined in $\bar{M}$. However, this does
not define a strong enough topology to give suitable open
neighbourhoods of all ideal points. We will therefore work with
topologies which are based more indirectly on the causal structure. We
face some of the same difficulties as earlier authors, and we do not
find a unique topology which yields intuitively appealing results in
all examples.  Instead, we will explore two different definitions of
the topology, representing different implementations of the same basic
idea.

We introduce a topology $\bar{\cal T}$ in section~\ref{top}. We
introduce a new structure on which to base the topology, using the
causal structure to define sets $L^\pm(\bar{S})$ which it is
appropriate to regard as the closure in $\bar{M}$ of
$I^\pm_{C}(\bar{S})$ for arbitrary $\bar{S} \subset \bar{M}$. We then
define a topology on $\bar{M}$ by requiring that the sets $\bar{M}
\setminus L^\pm(\bar{S})$ are open. We show that in this topology,
$\bar{M}$ does correspond to $M$ equipped with a boundary in the
topological sense, and that ideal point pairs containing
$I^\pm[\gamma]$ are limit points of the timelike curve $\gamma$. We
show that this topology has fairly good separation properties;
failure of the expected separation of ideal points occurs only in
well-controlled examples. On the other hand, some sequences of points
in $M$ will fail to have the intuitively expected limit points in
$\bar M$. This motivates us to consider a modified topology $\bar{\cal
T}_{alt}$ in appendix~\ref{app:weak}, in which we use a different
definition of the sets $L^\pm(\bar{S})$. This modified topology leads to more limit
points in interesting examples, 
but it goes a bit too far and has weaker
separation properties.  These two topologies share some desirable
features: both satisfy the minimal requirements enunciated above, and
they yield the standard topology on the completion for both de Sitter
and anti-de Sitter space. 

As was remarked in~\cite{Geroch}, one does not have any sense in the
abstract of what is the correct completion of a spacetime $M$. In
judging a proposed completion, applying our intuition to simple
examples is an important source of guidance. Such examples have played
an important part in the previous literature on this
subject~\cite{diff1,diff2,diff3}. We will use several simple toy
examples to illustrate the different aspects of the construction as we
go along. Once we have developed the techniques, we will apply them to
the construction of the causal boundary for plane waves.  Plane waves
provide an important `real-world' example where the full power of the
causal approach is needed to obtain an asymptotic boundary. In
section~\ref{plane} we find that, using these purely causal methods,
one may recover the result (previously announced in~\cite{beyond})
that the causal boundary of a homogeneous plane wave is a null line
segment. A careful application of the original definition of
\cite{Geroch} gives a different answer, illustrating the importance of
these subtleties.

Due to the length of time which has passed since the works
\cite{Geroch,budic,szab1} on which we build, we have tried to make
this paper reasonably self-contained.  Although we have assumed
familiarity with the basic ideas of causality commonly used in general
relativity, it is hoped that the interested reader having no previous
familiarity with the specifics of causal completions will be able to
understand our discussion without reference to the previous
literature.  We have not reviewed all the previously proposed
constructions in detail, but we have tried to state the differences in
their actions on the examples in a way that does not assume a
knowledge of these details.

\section{Construction of a causal completion $\bar M$}

Following previous works, we first assume that the spacetime is
strongly causal; this seems to be the weakest assumption allowing us
to construct a causal completion $\bar{M}$ in general. A strongly
causal spacetime is a spacetime in which, for every $x \in M$, there
is a neighborhood $U \ni x$ which no non-spacelike curve enters more
than once. Strong causality implies that the spacetime is past- and
future-distinguishing i.e., $I^-(p) = I^-(q)$ iff $p=q$, and $I^+(p) =
I^+(q)$ iff $p=q$, for any points $p,q \in M$.  This means that we can
associate any point $p$ in $M$ with its chronological past and future
$I^\pm(p)$ in a one-to-one way.

\subsection{IPs and IFs}
\label{rev}

The set $I^-(p)$ is an example of a past-set: a set which is the
past of some set. In fact, $I^-(p)$ is an indecomposable past-set, or
IP. This means it cannot be expressed as the union of two proper
subsets which are themselves past-sets. If we think of the spacetime
in terms of its causal structure, these indecomposable past-sets can
be thought of as the elementary objects from which the causal
structure is built.

\begin{figure}
\begin{center}
    \includegraphics[width=0.8\textwidth]{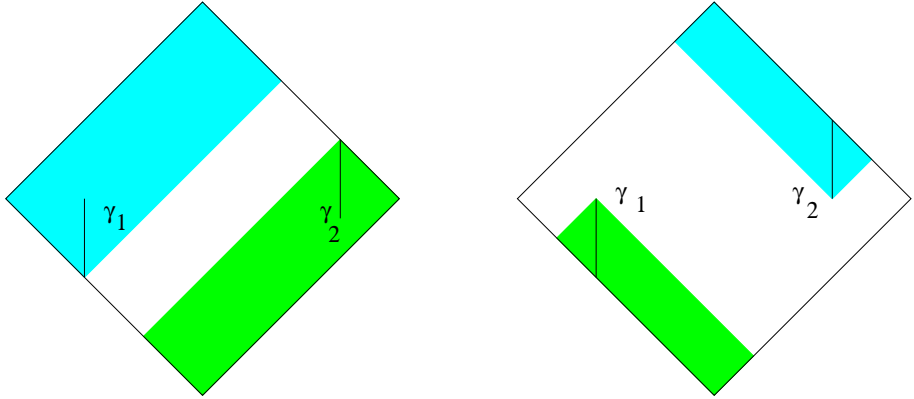}
\end{center}
\caption{On the left, a conformal diagram of 1+1 Minkowski space
showing the TIF $I^+[\gamma_1]$ and TIP $I^-[\gamma_2]$.  On the right, 
the PIP $I^-[\gamma_1]$ and PIP $I^-[\gamma_2]$ are shown.
}\label{fig:endpoints}
\end{figure}

So every point $p \in M$ determines an IP $I^-(p)$. However, the
converse is not true: there are IPs which are not of the form $I^-(p)$
for some $p \in M$. But Geroch, Kronheimer and Penrose
showed~\cite{Geroch} that every IP is of the form $I^-[\gamma]$ for
some timelike curve $\gamma$ in $M$, and that all such $I^-[\gamma]$ are IPs. 
Thus, the set of IPs is the set
of chronological pasts $I^-[\gamma]$ for all timelike curves $\gamma$
in $M$. If the curve $\gamma$ has a future endpoint $p$, $I^-[\gamma]
= I^-(p)$, the IP is termed a proper IP, or PIP. If $\gamma$ is
future-endless, the IP $I^-[\gamma]$ is a terminal IP, or TIP. We call
the set of IPs $\hat{M}$. Elements of this set are denoted by
undecorated capital letters, $P \in \hat{M}$.

One can similarly define future-sets, indecomposable future-sets
(IFs), PIFs and TIFs. The set of IFs is called $\check{M}$. Elements
of this set will be written with stars, $P^* \in \check{M}$.

The idea is to construct the ideal points from the TIPs and TIFs; that
is, we construct the additional points in terms of the spacetime
regions they can physically influence or be influenced by. Put another
way, we will add endpoints to endless timelike curves, in such a way
that two curves go to the same future (past) point if they have the
same past (future). Thus, we think of $\hat{M}$ as a partial
completion of $M$, which adjoins as ideal points the `future
endpoints' of all future-endless timelike curves $\gamma$. If the
spacetime $M$ is geodesically incomplete, the ideal points associated
with incomplete geodesics will represent singularities, so the ideal
points can represent both singularities and points at infinity.

Although this procedure does not give us as much structure as the
method of conformal rescaling, the fact that it is constructive is a
considerable advantage. Since we construct $\hat{M}$ directly from
$M$, this method is universally applicable.  In contrast, conformal
compactification makes the assumption that a suitable
conformally-related manifold $\tilde{M}$ exists.  This assumption can
fail, and even if such an $\tilde{M}$ exists, we have no general
algorithm for obtaining it.  Finally, although uniqueness of the conformal approach in the 
asymptotically simple context was shown in \cite{Penrose}, it need not be unique more generally.  
But in the causal approach the construction of
$\hat{M}$ is a straightforward exercise in studying the causal
structure, and the only assumption is strong causality. 
This method also allows us to study singular points on the
same footing as points at infinity, which is not possible in the
conformal compactification.

\subsection{Quotient constructions}
\label{iden}

\begin{figure}
\begin{center}
    \includegraphics[width=0.5\textwidth]{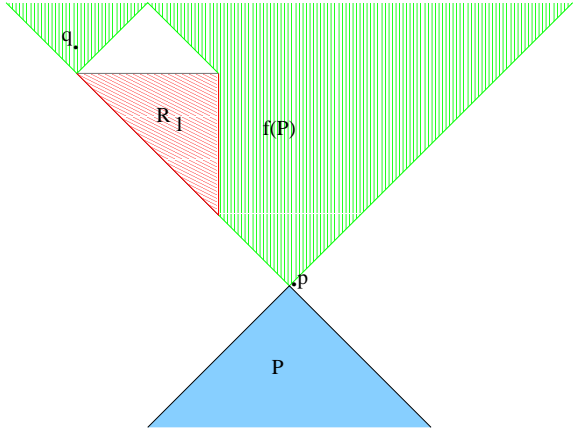}
\end{center}
\caption{An example which illustrates important subtleties in dealing
with the causal completion. The spacetime is constructed by excising the
closed triangular region $R_1$ from two-dimensional flat space. If we
consider a point $p$ on the diagonal null line, the PIP $P = I^-(p)$
provides an example where $f(P)$ is not an IF. Also, the points $p$
and $q$ are causally related in $\bar{M}$, but not in $M$.
}\label{fig:triangle}
\end{figure}

Clearly what we have constructed so far are only partial completions;
we should wish to adjoin both the past and future ideal points, that
is, both the TIPs and TIFs, in constructing our completion
$\bar{M}$. Since there is a natural mapping $M \to \{ \hat{M},
\check{M} \}$, conveniently referred to as $I^\mp$, which sends $p \to
\{ I^-(p), I^+(p) \}$, one might try to construct a completion which
adjoins all TIPs and TIFs by considering the set
\begin{equation} \label{quot1}
M^\sharp = (\hat{M} \cup \check{M})/R,
\end{equation}
where $R$ identifies every PIP $I^-(p)$ with the corresponding PIF
$I^+(p)$. Elements of $M^\sharp$ will be written as $P^\sharp$.
However, this will overcount the ideal points; in some cases, there
will be timelike curves in $M$ which approach a given ideal point from
both the past and the future (this happens, for example, for the
conformal boundary of anti-de Sitter space, or on the vertical segment
of the triangle $R_1$ in the example of figure~\ref{fig:triangle}). In
an approach based on identifications, some TIPs and TIFs must also be
identified.

In~\cite{Geroch}, additional identifications between TIPs and TIFs
were obtained by introducing a topology on the $M^\sharp$ defined
above and making further identifications as required to obtain a
Hausdorff space. Apart from practical problems with the specific
implementation in~\cite{Geroch} which have been extensively discussed
in the literature~\cite{diff1,diff2,racz,szab1,diff3}, this
topological approach has two conceptual defects: it makes the
connection between the identifications and the causal structure rather
indirect, and it introduces an additional assumption, that the correct
completion will be Hausdorff in this topology. Although this might at
first seem an entirely reasonable assumption, there is no natural
reason to impose any such a priori restriction on the completion for
arbitrarily complicated $M$.  In fact, an example (see figure
\ref{fig:top}) was introduced in \cite{Geroch} for which $\bar M$
cannot simultaneously be Hausdorff, have the intuitively expected limiting ideal points for sequences in $M$, and have a chronology compatible with that of $M$.

As a result, Budic and Sachs \cite{budic} proposed to construct the
identifications between TIPs and TIFs in a different way. They modified
the quotient construction (\ref{quot1}) by finding an appropriate
identification $R$ which can act on any IPs and IFs, and whose action
on PIPs and PIFs is equivalent to that used in (\ref{quot1}). One
wishes to identify IPs and IFs if they are the past and future of `the
same point'. Thus, one regards the identification used in
(\ref{quot1}) as a two-step process, first introducing a notion of the
`future' of a PIP $P = I^-(p)$ as the set $f(P) = I^+(p)$, and
similarly defining the `past' of a PIF $P^* = I^+(p)$ as the set
$p(P^*) = I^-(p)$, and then identifying an IP $P$ with an IF $Q^*$
if $f(P)$ is appropriately related to $Q^*$ and $p(Q^*)$ is
appropriately related to $P$.

Given a past-set $P$, a natural notion of the associated future-set is
the `common future', that is, the set of points to the future of all
points $p \in P$. More formally, define $f(P)$ for any IP $P$ by
\begin{defn}
The common future $f(P)$ for any IP $P$ is
\begin{equation}
f(P) =  \cup_x I^+(x) : x \in M,  P \subset I^-(x).
\end{equation}
\end{defn}
That is, $f(P)$ is the union of the futures of all points $x \in M$
such that $x$ is to the future of all points in $P$. This can be
rewritten in several equivalent ways:
\begin{eqnarray}
f(P) &=& I^+[ \{ x \in M: P \subset I^-(x) \}] \\
&=& I^+[ \cap I^+(p): p \in P] \\
&=& Int[  \cap I^+(p): p \in P],
\end{eqnarray}
where $Int$ denotes the topological interior of a set in $M$.
The set $f(P)$ so defined is by definition a future-set; however, it
is not necessarily an {\it indecomposable} future-set (see
figure~\ref{fig:triangle} for a counterexample).  The common past
$p(P^*)$ of the future set $P^*$ is defined similarly.

In~\cite{budic}, the spacetime $M$ was restricted to be causally
continuous, which means that $I^\pm(p)$ will vary continuously (as a
point-set in $M$) whenever $p$ does. This restriction was meant to
eliminate all the examples where $f(P)$ is not an IF (in fact, it does
not quite do so; see~\cite{rube} for a counterexample),
and~\cite{budic} then basically defined $R$ to identify $P$ with $Q^*$
iff $f(P) = Q^*$ and $P = p(Q^*)$.\footnote{In fact, the completion was
technically defined in~\cite{budic} by deleting certain points from
$\hat M \cup \check M$ rather than by making identifications. However,
in this causally continuous context, the definition used
in~\cite{budic} is equivalent to making identifications.}

This approach was extended by Szabados~\cite{szab1}, who introduced a
more subtle identification which successfully addresses examples such
as figure~\ref{fig:triangle}. We present his idea in two stages.
First consider the following relation
\begin{defn} \label{rpf}
Let $R_{pf} \subset \hat{M} \times \check{M}$ be the set of pairs of
the form $(P,Q^*)$ for which $Q^*$ is a maximal subset of $f(P)$ and
$P$ is a maximal subset of $p(Q^*)$; i.e., satisfying
\begin{equation} \label{cond1}
Q^* \subset f(P); \quad \nexists R^* \in \check{M}: R^* \neq Q^*, Q^*
\subset R^* \subset f(P).
\end{equation}
and
\begin{equation} \label{cond2}
P \subset p(Q^*); \quad \nexists R \in \hat{M}: R \neq P, P \subset R
\subset p(Q^*).
\end{equation}
\end{defn}
Now consider $\bar R_{pf}$, the smallest equivalence relation on
$\hat{M} \cup \check{M}$ which identifies every IP $P$ and IF $Q^*$
for which $(P,Q^*) \in R_{pf}$.  Szabados' idea \cite{szab1} is then 
equivalent to\footnote{Szabados also introduced a further
identification directly between pairs of TIPs or TIFs, in
\cite{szab2}. This was necessary to achieve what were considered to be
satisfactory separation properties in the topology used
in~\cite{szab2}. In our approach, there will be no analogue of these
additional identifications in the construction of $\bar M$. The
construction of quotients of $\bar M$ with various separation
properties will be discussed in more detail in section \ref{sep}.}
the following identification:
\begin{defn}
Given $M$, the causal completion $\bar{M}_{id}$ is
\begin{equation} \label{quot2}
\bar{M}_{id} = (\hat{M} \cup \check{M})/\bar R_{pf}.
\end{equation}
\end{defn}
Note that if $f(P)$ is an IF, the above condition reduces to $f(P) =
Q^*$, in which case they are the same as those
of~\cite{budic}. The subscript $id$ refers to the construction through the
identifications imposed on points in $\hat{M} \cup \check{M}$.  An important
theorem proven in \cite{szab1} (his Proposition 5.1) is
\begin{thm}
\label{SzabThm}
For each point $q \in M$, the identification $\bar R_{pf}$ used in the
preceding definition 
identifies $I^+(q)$with $I^-(q)$ but with no other IPs or IFs.
\end{thm}

\begin{figure}
\begin{center}
    \includegraphics[width=0.8\textwidth]{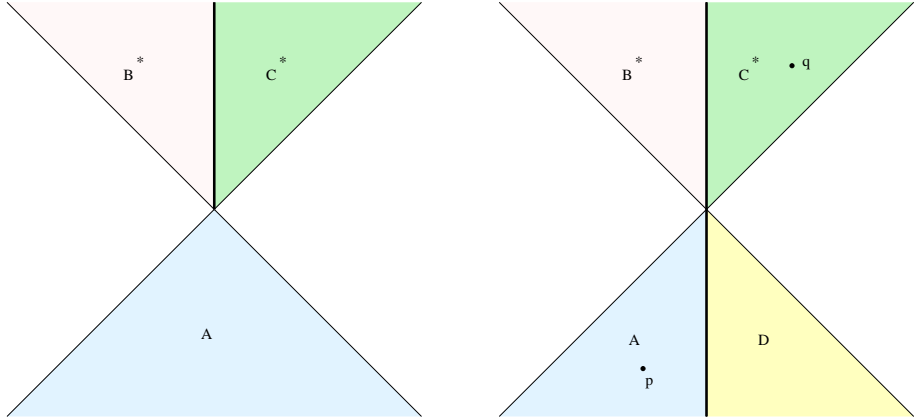}
\end{center}
\caption{An example (left) where more than one TIF is identified with
  a TIP. We excise the line $x=0$, $t \geq 0$ from two-dimensional
  flat space. The TIP $A$ and TIFs $B^*$, $C^*$ all correspond to the
  same ideal point. A trouble-free but suggestive example in which we
  consider the IPs and IFs in the interior of an excised line segment
  appears on the right.}\label{fig:two}
\end{figure}

Thus, the identification $\bar R_{pf}$ will reproduce precisely the
identification $R$ used to construct $M^\sharp$ in (\ref{quot1}),
together with some additional identifications that involve only TIPs
and TIFs. This property is a good start. Note, however, that while
each PIP is identified with exactly one PIF and vice-versa, the same
need not apply to TIPs and TIFs; the left spacetime in
figure~\ref{fig:two} illustrates an example where more than one TIF is
identified with a TIP. This property turns out to be quite
dangerous. It implies that the future of an ideal point in this
approach is not always an IF; as illustrated by the 2+1 dimensional
example in appendix \ref{app:ex}, the addition of ideal points can
introduce new causal relations between elements of $M$.

While the 2+1 dimensional example in appendix \ref{app:ex} is not
overly complicated, we know of no useful 1+1 analogue.  Thus it cannot
be summarized by a simple picture and its description is somewhat
involved.  However, the basic flavor can be obtained by consulting
figure \ref{fig:two}.  In the example on the left, the pairs $(A,B^*),
(A,C^*) \in R_{pf}$ force the identification of $B^*$ and $C^*$,
though these two future sets have little in common.  In contrast the
example on the right has only $(A,B^*),(D,C^*) \in R_{pf}$, and $B^*$
is not identified with $C^*$.  Note, however, that if for some reason
$B^*$ and $C^*$ were identified then a causal link would be introduced
between $p$ and $q$ even though $q \notin I^+(p)$; the new ideal point
would naturally be thought of as lying chronologically between $p$ and
$q$ and forming a new causal link.  The 2+1 dimensional example in
appendix \ref{app:ex} achieves essentially this result through the use
of a fifth region $E$.  In sufficiently complicated examples the
introduction of such new causal links can create closed timelike
curves in the completed manifold. We find this property
unsatisfactory, as it destroys the very structure on which the causal
completion was built.

\subsection{Ideal points as IP-IF pairs}
\label{pairs}

Our main purpose in this paper is to introduce a new approach to the
construction of the causal completion $\bar M$ from the primitive data
about the causal structure of the spacetime encoded in $\hat M$ and
$\check M$. The basic idea is that we want to get back to thinking of
the TIPs and TIFs as the pasts and futures of ideal points. That is,
we want to define elements of $\bar M$ in such a way that the past (in
$M$) of any element of $\bar M$ is either some element of $\hat M$, or
is empty.\footnote{Note that we must allow the latter possibility to address
ideal points that are in the `past boundary' of $M$; these must then
have a non-empty future.} As we have seen above, this will not be true
in an approach based on equivalence classes when one considers a
reasonably general class of spacetimes including, e.g., the left
example in figure \ref{fig:two}.

As a result, we advocate a different construction. We will define an
element $\bar P$ of $\bar M$ in terms of an IP $P$ and an IF $P^*$.
Later, we will interpret $P$ and $P^*$ as the intersection with $M$ of
the past and future of $\bar P$.  We adopt the same relation $R_{pf}$
defined in definition~\ref{rpf} to relate pasts and futures, but use
$R_{pf}$ directly instead of completing it to an equivalence relation
$\bar R_{pf}$:
\begin{defn} \label{compl}
Given $M$, the causal completion $\bar{M}$ consists of all those pairs
$(P,P^*)$ such that
\begin{enumerate}[i)]
\item $(P,P^*) \in R_{pf} \subset \hat M \times
\check M$, or
\item   $P = \emptyset$ and $P^*$ does not appear
in any pair in $R_{pf}$, or
\item   $P^* = \emptyset$ and $P$ does not appear
in any pair in $R_{pf}$.
\end{enumerate}
\end{defn}
Note that every IP $P$ or IF $P^*$ appears in at least one pair; they
may appear in more than one such pair, though those that appear in
pairs of the form $(P,\emptyset)$ or $(\emptyset,P^*)$ appear in only
one pair.

Elements of $\bar{M}$ will be written as $\bar{P} = (P,P^*)$. In
discussing some element $\bar{P}$, we will write $P,P^*$ to mean the
elements of the pair.  Note that theorem \ref{SzabThm} guarantees
that, for $p \in M$, the sets $I^\pm(p)$ each appear in exactly one
pair in $\bar M$, and that they appear together.  Thus there is a
natural injective map $\Phi: M \to \bar M$, $\Phi(p) = (I^-(p),I^+(p))$. 

In the example on the left of figure~\ref{fig:two}, instead of having
a single ideal point whose past is $A$ and whose future is $B^* \cup
C^*$, we will have two ideal points $(A, B^*)$ and $(A, C^*)$. We feel
that this is a more natural result from the point of view of the
causal structure; there is nothing about the causal structure that
requires these two points be identified---there is no relation between
$B^*$ and $C^*$---and it is useful to record the information about the
separate relations between $A$ and $B^*$ and $A$ and $C^*$ in the
causal completion. If one feels that consideration of the causal
structure alone produces a rather `big' completion, the resulting
$\bar M$ can always be used as a starting-point for further
identifications, should they seem desirable for some other reason.

\section{Causal structure on $\bar{M}$}
\label{struct}

Since the causal structure is of primary importance, one would like to
see to what extent the causal structure on $M$ extends to $\bar{M}$. A
general axiomatic definition of a causal space was given
in~\cite{kpcausal}:
\begin{defn} A set $M$ together with three relations $\prec$,
$\ll$ and $\to$ is called a causal space if the relations satisfy the
following conditions:
\begin{enumerate}
\item $x \prec x$,
\item $x \prec y$ and $y \prec z$ implies $x \prec z$,
\item $x \prec y$ and $y \prec x$ implies $x = y$,
\item Not $x \ll x$,
\item $x \prec y$ and $y \ll z$ implies $x \ll z$,
\item $x \ll y$ and $y \prec z$ implies $x \ll z$,
\item $x \to y$ iff $x \prec y$ and not $x \ll y$.
\end{enumerate} \label{causdefn}
\end{defn}
The relations are collectively called the causal relations; $\prec$ is
called the causality, $\ll$ is called the chronology, and $\to$ is
called the horismos. If we have two relations satisfying the relevant
conditions, we can use them to define the third; for example, we can
regard the last condition as defining the horismos.  

Before discussing the causal structure for $\bar{M}$, let us review
some results of~\cite{Geroch}, where it was shown that we can make the
set $\hat{M}$ a causal space by introducing a causality and chronology
specified by
\begin{equation}
P \prec_{IP} Q \quad \mbox{iff } P \subset Q,
\end{equation}
\begin{equation} \label{ipchron}
P \ll_{IP} Q \quad \mbox{iff } P \subset I^-(q) \mbox{ for some } q \in Q.
\end{equation}
They showed that when $P$ and $Q$ are PIPs, these relations include
the relations between $p$ and $q$ as points of $M$. That is, $p \prec
q$ implies $I^-(p) \prec_{IP} I^-(q)$, and $p \ll q$ implies $I^-(p)
\ll_{IP} I^-(q)$. One says that the map $I^-$ preserves the chronology
and the causality.  However, $p$ and $q$ in figure \ref{fig:triangle}
show that the inverse statements do not hold.  There, $p \prec_{IP} q$
since $I^-(p) \subset I^-(q)$, but we do not have $p \prec q$.  In
addition, $p \ll_{IP} q$ but we do not have $p \ll q$.  Thus, this
causality and chronology can involve additional relations even between
PIPs. This suggests that it is necessary to consider both pasts and
futures to construct a good chronology. 

We can similarly define causality and chronology on $\check{M}$, using
  $Q \prec_{IF} P$ if $P \subset Q$, $Q \ll_{IF} P$ if $P \subset
  I^+(q)$ for some $q \in Q$.

\subsection{Chronology on $\bar{M}$}
\label{chron}

Let us now proceed to consider $\bar{M}$. There is a natural
chronology on $\bar M$:
\begin{thm} The relation $\ll_C$ on $\bar{M}$ defined by 
\label{ChronThm}
\begin{equation} \label{barchron}
\bar{P} \ll_C \bar{Q} \quad \mbox{iff }  P^* \cap Q \neq \emptyset.
\end{equation}
is a chronology on $\bar M$. 
\end{thm}
{\it Proof:} To show transitivity, suppose $\bar{P} \ll_C \bar{Q}$ and
$\bar{Q} \ll_C \bar{R}$. Since $\bar{P} \ll_C \bar{Q}$, there is an $s
\in P^* \cap Q$. Since $\bar{Q} \ll_C \bar{R}$, there is a $t \in Q^*
\cap R$. Since $t \in Q^*$, $t \in f(Q)$, so $s \in I^-(t)$. But $t
\in R$ implies $I^-(t) \subset R$, so $s \in R$, and $P^* \cap R \neq
\emptyset$, implying $\bar{P} \ll_C \bar{R}$. 

We prove that it is anti-reflexive by assuming the converse. Imagine
there was a $\bar{P} \in \bar{M}$ such that $\bar{P} \ll_C \bar{P}$:
this implies $P^* \cap P \neq \emptyset$. Consider $x \in P^* \cap P$;
since $P$ and $P^*$ were related, $P \subset p(P^*)$, implying $P
\subset I^-(x)$, and hence $x \in I^-(x)$, in contradiction to the
assumption that $M$ is a causal space. Thus, $\ll_C$ is an
anti-reflexive partial ordering, and hence a chronology. $\Box$

That this chronology is preferred over one induced from $\ll_{IP}$ can
be seen from the example in figure~\ref{fig:triangle}.  While $I^-(p)
\ll_{IP} I^-(q)$, $p$ and $q$ are not chronologically related by
$\ll_C$.  Thus, the chronology $\ll_C$ better preserves the chronology
of $M$.\footnote{Note that the chronology relation (\ref{ipchron}) can
be re-expressed as $P \ll_{IP} Q$ iff $f(P) \cap Q \neq \emptyset$,
showing why we can define a more satisfactory chronology on $\bar{M}$;
the IP $P^*$ that $P$ is paired with provides a better notion of
`the future of $P$' than $f(P)$.}

The relation $\ll_C$ is directly analogous to $\bar{\ll}$ on $\bar
M_{id}$ introduced in \cite{szab1}.  However, when applied to the $\bar M_{id}$
associated with the example in appendix \ref{app:ex}, $\ll_C$ is not transitive
as defined in \ref{barchron}.  Theorem \ref{ChronThm} shows that this
cannot happen on our $\bar M$.

In our nomenclature, the subscript $C$ stands for `curve', as
illustrated by the following result:
\begin{thm}
\label{chronmatch}
Two points are chronologically related, $\bar{P} \ll_C \bar{Q}$, if
and only if there exists a timelike curve $\gamma_{PQ}$ such that $P^*
= I^+[\gamma_{PQ}]$ and $Q= I^-[\gamma_{PQ}]$. \label{lemc}
\end{thm} 
{\it Proof:} If $P^* =I^+[\gamma_{PQ}]$, $Q=
I^-[\gamma_{PQ}]$, clearly $P^* \cap Q \neq \emptyset$. If $P^* \cap Q
\neq \emptyset$, consider a point $r \in P^* \cap Q$, a curve
$\gamma_1$ such that $P^* = I^+[\gamma_1]$, and a curve $\gamma_2$
such that $Q = I^-[\gamma_2]$. Then by definition, $s \ll r$ for some
point $s$ on $\gamma_1$, and $r \ll t$ for some point $t$ on
$\gamma_2$. We can then construct the required curve $\gamma_{PQ}$ by
joining together the portion of $\gamma_1$ before $s$, a timelike
curve from $s$ to $r$, a timelike curve from $r$ to $t$, and the
portion of $\gamma_2$ after $t$. $\Box$

This shows that $\ll_C$ is a natural chronology--- $\bar{P} \ll_C
\bar{Q}$ if and only if there is a timelike curve `from $\bar{P}$ to
$\bar{Q}$'.

An immediate consequence of the above theorem is
\begin{corr}
On $M$, $\ll$ is equivalent to the chronology induced on $M$ by the
natural embedding in $\bar M$ with chronology $\ll_C$.
\end{corr} 
That is, $M$ is chronologically isomorphic to its image in $\bar M$.
Although~\cite{budic} proves a
similar theorem for the chronology $\ll_{IP}$, it is valid only in
much more restrictive circumstances. This exhibits why the proposed
chronology (\ref{barchron}) is preferred.   

Furthermore, we have the
following useful result:

\begin{thm}
The chronology $\ll_C$ defined by (\ref{barchron}) is weakly
distinguishing: $I^+_C(\bar{P}) = I^+_C(\bar{Q})$
and $I^-_C(\bar{P}) = I^-_C(\bar{Q})$ if and only if $\bar{P} =
\bar{Q}$. 
\end{thm}
{\it Proof:} If we consider a point $\bar{P} = (P,P^*) \in \bar M$,
the set of points associated with PIPs in $I^+_C(\bar P)$ is
clearly in one-one correspondence with the points in the IF $P^*$,
which may be empty: $P^* = I^+_C(\bar{P}) \cap M$ . 
Hence $I^+_C(\bar{P}) = I^+_C(\bar{Q})$ for two
points in $\bar M$ implies $P^* = Q^*$.  Similarly, $I^-_C(\bar{P}) =
I^-_C(\bar{Q})$ implies $P = Q$.  Thus, if both $I^\pm_C(\bar P)$ and
$I^\pm_C(\bar Q)$ agree we must have $(P,P^*) = (Q,Q^*)$.  The converse
is trivial.  $\Box$

\subsection{Causality on $\bar{M}$}
\label{caus}

\begin{figure}
\begin{center}
    \includegraphics[width=0.5\textwidth]{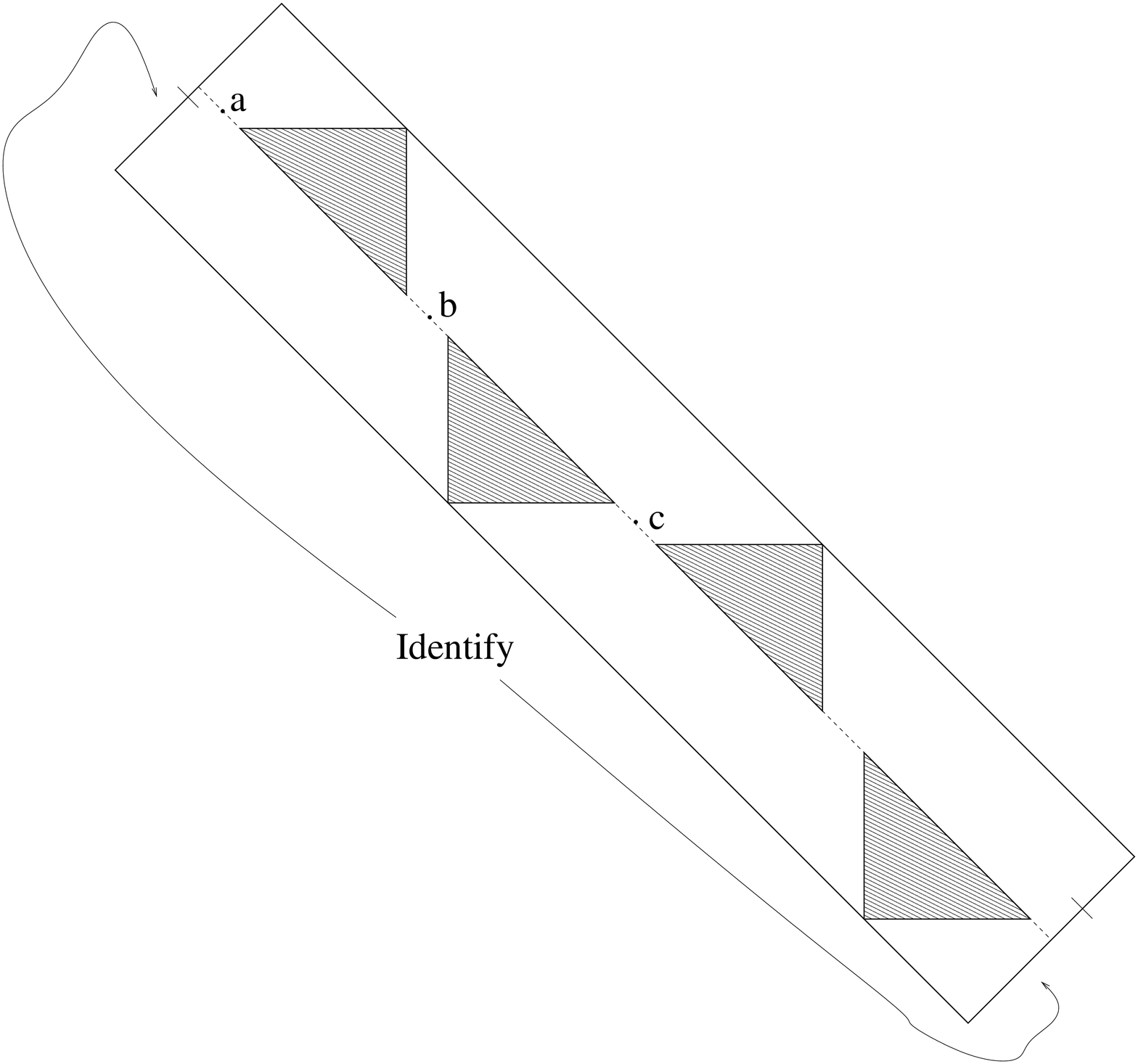}
\end{center}
\caption{An illustration of the problems with constructing causality
  on $\bar{M}$. The spacetime is the interior of the rectangle, with
  the edges marked identified and the triangular regions removed; this
  spacetime is strongly causal. If we define a causality on $\bar{M}$
  by $\bar{P} \prec_I \bar{Q}$ if $P \prec_{IP} Q$ or $P^* \prec_{IF}
  Q^*$, it will not give a partial ordering; $c \prec_I b$ and $b
  \prec_I a$, but $c \nprec_I a$. If we add additional causal
  relations to make a partial ordering, we will have $c \prec_I a$ and
  $a \prec_I c$, and the dashed line will be a closed causal curve.
  However, one may also choose to introduce the causality defined by
  the chronology in (\protect\ref{ccaus}) so that $c \nprec_C b$.  Thus, if
  one is willing to pay the price of losing the causal link between
  $b$ and $c$, one can eliminate closed causal curves in $\bar M$.}
\label{fig:badcaus}
\end{figure}

The situation is somewhat less satisfactory as concerns the causality.
There is no natural way to directly define a causality on
$\bar{M}$. Thus, we must define it indirectly. One approach is to
define it using the causalities on $\hat{M}$ and $\check{M}$. The
difficulty here is that if we define a causality by $\bar{P} \prec_I
\bar{Q}$ if $P \prec_{IP} Q$ or $P^* \prec_{IF} Q^*$, it will fail on
several counts.  First, in the left example of figure \ref{fig:two},
it fails to satisfy criterion 3: we have both $(A,B^*) \prec_I
(A,C^*)$ and $(A,C^*) \prec_I (A,B^*)$.  In addition, $\prec_I$ fails
to be a partial ordering in figure \ref{fig:badcaus}.  Here we can
define a candidate causality by saying $\bar{P} \prec_I \bar{Q}$ if
there exists a chain of points $\bar{R}_i$, $i=1,\ldots,n$, such that
$P \prec_{IP} R_1$ or $P^* \prec_{IF} R_1^*$, $R_i \prec_{IP} R_{i+1}$
or $R^*_i \prec_{IF} R^*_{i+1}$ for each $i$, and $R_n \prec_{IP} Q$
or $R_n^* \prec_{IF} Q^*$. However, once again, we have been forced to
consider a transitive closure.  Figure~\ref{fig:badcaus} demonstrates
that in general this definition introduces additional causal
relationships between points of $M$, so that we cannot hope that $M$ will
be causally isomorphic to its image in $\bar M$.  We again see that
closed causal curves may result; that is, $\prec_I$ may fail to satisfy
property 3 in the definition of a causality. Furthermore, this
causality will generally fail to satisfy conditions 5 and 6 of
definition~\ref{causdefn} with respect to the chronology $\ll_C$.

Alternatively, we can define the causality from the chronology. Since
$\ll_C$ is a weakly distinguishing chronology, it determines a
causality (see~\cite{kpcausal} for the proof). That is, one may define
\begin{equation} \label{ccaus}
\bar{P} \prec_{C} \bar{Q}  \quad \mbox{iff } I^-_C(\bar{P}) \subset
I^-_C(\bar{Q}) \mbox{ and } I^+_C(\bar{P}) \supset I^+_C(\bar{Q}). 
\end{equation}
The causality $\prec_C$ satisfies conditions 5 and 6 of
definition~\ref{causdefn}, and hence $\bar M$ with chronology $\ll_C$
and causality $\prec_C$ is a causal space (essentially by
construction). In \cite{szab1}, this causality was briefly described
and shown to define a causally simple $\bar{M}_{id}$.  Thus, from a
formal point of view, this causality is much better behaved than
$\prec_I$.\footnote{However, $M$ can be causally isomorphic to its
image in $\bar M$ with the causality $\prec_C$ only if we require that the
causality $\prec$ on $M$ agrees with the analogue of $\prec_C$; that
is, we must require $p \prec q$ iff $I^-(p) \subset I^-(q)$ and $I^+(p) \supset
I^+(q)$. Such a $M$ is called a $\cal B$-space in the terminology
of~\cite{kpcausal}. This is a strong requirement; for example, flat
space with one point removed is not a $\cal B$-space.}
Note that  $\bar P
\prec_C \bar{Q}$ implies $\bar P \prec_I \bar Q$, the converses are
not generally true; for example, in figure~\ref{fig:badcaus}, $c
\prec_I b$, but $c \nprec_C b$.

However, in practice, this causality fails to give the expected causal
relations in some simple examples. Again we refer to
figure~\ref{fig:badcaus}, where one would expect $c \prec b$, but one
finds $c \nprec_C b$.  Another example, discussed in~\cite{diff2},
shows that $\prec_C$ yields unexpected results even for causally
continuous spaces.

Our approach to this issue is to admit to a certain ambivalence with
respect to causality.  Our feeling is that physical effects connected
with signal propagation are related to the chronology $I^\pm_C(\bar
P)$ and perhaps to the closure of such sets.  This closure is
determined by the topology, which will be discussed in section
\ref{top} below, but again figure \ref{fig:badcaus} shows that such
relations are unlikely to lead to a causality.  Thus, any desire to
impose a causality on $\bar M$ must stem from other motivations that
may vary with the precise context.  If one insists on making $\bar{M}$
a causal space, one can do so using $\prec_C$.  If, on the other hand,
one finds it useful to encode apparent null relations such as those
linking $a,b,c$ in figure \ref{fig:badcaus}, one may use $\prec_I$.
In general, however, it is not possible to do both. 

\section{Topology on $\bar{M}$}
\label{top}

The causal structure discussed in the previous section yields some
information about the relation between the causal boundary and points
in the interior. However, to provide a more complete description, we
need to make $\bar{M}$ into a topological space. In the quotient
constructions, the topology on $\bar{M}$ has played a central role
in determining appropriate identifications to form the quotients, and
it has been natural to construct the topology by the quotient of some
suitable topology on the space of all IPs and IFs.

In this section, we will propose a new definition for the topology on
the completion $\bar{M}$ we defined in section~\ref{pairs}. We must
construct our topology directly on $\bar{M}$. Following the philosophy
of the previous sections, it is directly based on the causal structure
of $\bar{M}$. In section~\ref{emb}, we will show that with this
topology, the map $M \to \bar{M}$ is an open dense embedding, so
$\bar{M}$ does correspond to $M$ equipped with boundary points in a
suitable topological sense.  In section~\ref{sep}, we consider the
separation properties of our topology. As in~\cite{Geroch}, we will
often omit dual results, obtained by exchanging future and past.

However, our definition of the topology will be fairly technical, and
it is interesting to explore to what extent its features depend on the
particular choices made in the definition. We therefore define a
contrasting topology ${\bar {\cal T}}_{alt}$ by making slightly
different technical choices in appendix \ref{app:weak}.  We find that
this alternative shares many features with our main definition of the
topology, but has non-trivial differences in the treatment of some
limits and in its separation properties. Both definitions suffer from
(to some extent complementary) disadvantages, so we do not regard our
work as completely settling the issue of the topology. It does however
provide two data points which provide partial solutions, illustrate
many of the relevant issues, and may be useful for future
investigations.

Let us begin by reviewing previous work on the topology; although we
cannot use it directly, it is an important source of inspiration for
our definitions. In~\cite{Geroch}, a topology was first introduced on
the set $M^\sharp$ defined in (\ref{quot1}), and identifications were
then imposed to define a set $\bar{M}_{id}$ which was Hausdorff in the
induced topology. The topology on $M^\sharp$ was defined to be the
coarsest topology such that for each $A \in \check{M}$, $B \in
\hat{M}$, all the sets $A^{\rm int}$, $A^{\rm ext}$, $B^{\rm int}$ and
$B^{\rm ext}$ are open, where
\begin{equation}
\label{GKPint}
A^{\rm int} = \{P^\sharp: P \in \hat{M} \mbox{ and } P \cap A \neq \emptyset
\},
\end{equation}
\begin{equation}
\label{GKPext}
A^{\rm ext} = \{ P^\sharp: P \in \hat{M} \mbox{ and } \forall S
\subset M, P = I^-[S] \Rightarrow \mbox{ not } I^+[S] \subset A \},
\end{equation}
with $B^{\rm int}$ and $B^{\rm ext}$ defined similarly. These sets are
meant to correspond roughly to $I^-(A^\sharp)$, $M^\sharp \setminus
J^-(A^\sharp)$, $I^+(B^\sharp)$, and $M^\sharp \setminus
J^+(B^\sharp)$ respectively.  Thus, this topology is meant to
correspond to a generalised Alexandrov topology; the additional open
sets defined by the complement of the causal pasts and futures are
necessary to construct suitable open neighbourhoods of TIPs and TIFs,
as explained in~\cite{Geroch}.

\begin{figure}
\begin{center}
    \includegraphics[width=0.3\textwidth]{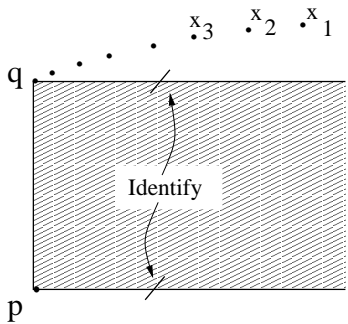}
\end{center}
\caption{A spacetime introduced in \protect\cite{Geroch} for which the
causal completion can be defined to have at most two of the following
three properties: i) causality, ii) Hausdorff topology, iii) the
convergence of $x_n$ to $p$ and $q$.  The shaded region is removed
from two-dimensional flat space, and edges indicated identified.  If
the sequence $x_n$ has either $p$ or $q$ as a limit point, it must
have both in any topology that does not distinguish past from future.
In our causal construction, $p$ and $q$ are not identified.  In the
topology of definition \protect\ref{strongtop}, $\bar M$ is causal but
neither of the other two properties hold.  We introduce a contrasting
topology in appendix \protect\ref{app:weak} in which $x_n \rightarrow
p,q$ and $\bar M$ is causal but not Hausdorff. }  \label{fig:top}
\end{figure}

Although~\cite{szab1} uses the causal identification rule discussed
previously, the topology on $\bar{M}_{id}$ is still obtained from the
above topology on $M^\sharp$.  Various problems with this topology
have been discussed in the literature~\cite{diff1,diff2,diff3}. In
particular, in~\cite{diff3}, an example was constructed where certain
sequences of points in $M$ do not have the expected limit points in
$\bar{M}_{id}$. We will discuss a directly analogous example in the
context of plane waves in section \ref{plane}.  An attempt to improve
the GKP topology was made in~\cite{racz}, but it was also found to
have difficulties~\cite{diff3}. 

In~\cite{budic}, by contrast, the topology is defined directly on
$\bar M$. The topology defined in~\cite{budic} is a generalised
Alexandrov topology determined by the causal structures introduced on
$\bar M$. That is, the sets $A^{\rm int}$, $A^{\rm ext}$, $B^{\rm
int}$ and $B^{\rm ext}$ used in the GKP topology are replaced by the
$I^-(\bar A)$, $\bar M \setminus J^-(\bar A)$, $I^+(\bar B)$, $\bar M
\setminus J^+(\bar B)$ defined for elements $\bar A,\bar B \in \bar M$ by a suitable chronology and causality relations
on $\bar M$.   We will not discuss these relations in detail as they were intended to apply only to causally continuous spaces.

Let us now turn to the construction of a topology on our $\bar M$.
Since our whole approach centres around the definition of a chronology
$\ll_C$ on $\bar M$, the most natural topology is an Alexandrov
topology, where the chronological future- and past-sets
$I^\pm_C(\bar{Q})$ form a subbasis. However, as argued
in~\cite{Geroch}, this is not a sufficiently strong topology; it does
not provide suitable open neighbourhoods for all ideal points. One
might consider extending it by introducing a generalised Alexandrov
topology using both the chronology and the causality, as
in~\cite{budic}. This avenue is obstructed by the absence of a truly
satisfactory causality in $\bar M$ for general strongly causal
$M$. The only candidate in general is $\prec_C$; but in
figure~\ref{fig:badcaus}, $c \nprec b$, so $b \in \bar M \setminus
J^+_C(c)$, and $[\bar M \setminus J^+_C(c)] \cap M$ is not an open set
in the manifold topology on $M$. That is, $M$ is not homeomorphic to
its image in $\bar M$ with such a topology, so this topology is not
generally satisfactory.

We therefore need to adopt a more indirect route in the definition of
our topology.  However, we will see in Lemma~\ref{IC} below that the
topology we define will contain the Alexandrov topology; that is, all
the chronological future- and past-sets $I^\pm_C(\bar{Q})$ are open in
our topology. The topology we adopt is based on defining suitable
closed sets, in a modified version of the other part of the GKP
topology.  We want to use the chronology to define sets
$L^\pm(\bar{S})$ for any subset $\bar{S} \subset \bar{M}$, such that
we can define a reasonable topology by requiring $\bar{M} \setminus
L^\pm(\bar{S})$ to be open. We start by defining the sets
\begin{equation}
L^+_{IF}(\bar S) = \{ \bar Q \in \bar M : Q^* \neq \emptyset, Q^*
\subset \cup_{\bar P \in \bar S} P^* \},
\end{equation}
\begin{equation}
L^-_{IP}(\bar S) = \{ \bar Q \in \bar M : Q \neq \emptyset, Q \subset  
\cup_{\bar P \in \bar S} P \}.
\end{equation}
Note that for $p \in M$, $p \in \bar{S}$ implies $p \in
L^+_{IF}(\bar{S})$ and $p \in L^-_{IP}(\bar{S})$, as $I^+(p) \subset
\cup_{\bar P \in \bar S} P^*$ and $I^-(p) \subset \cup_{\bar P \in
\bar S} P$.  If we applied these definitions to a manifold, they would
give $Cl[I^\pm(\bar{S})]$, so this seems a natural way to define
closed sets from the causality.\footnote{Note that even applied to a
manifold, these need not agree with the causal future and past
$J^\pm(S)$, as the latter are not necessarily closed sets; for
example, in figure~\ref{fig:badcaus}, $b \in L^+_{IF}(c)$, while $b \notin
J^+(c)$.  Our $L^+_{IF}(p)$
also differs from the closed set $K^+(p)$ introduced in \cite{SorkW}, as
in figure~\ref{fig:triangle}, $q \in K^+(p)$ but $q \notin L^+_{IF}(p)$.}
However, as they stand, these cannot in
general be good closed sets in $\bar{M}$, because they only contain
IFs (respectively, IPs), and so can never contain TIPs (TIFs)
corresponding to future (past) boundary points. To rectify this, we
introduce operations $Cl_{FB}$ and $Cl_{PB}$, the closures in the
future and past boundaries, by
\begin{equation}
Cl_{FB}(\bar S) = \bar S \cup \{ \bar Q \in \bar M : Q^* = \emptyset,
Q = \lim \  P_n \ {\rm for} \ \{P_n\} \in \bar S \},
\end{equation}
\begin{equation}
Cl_{PB}(\bar
S) = \bar S \cup \{ \bar Q \in \bar M : Q = \emptyset, Q^* = \lim \ 
P_n^* \ {\rm for} \ \{P^*_n\} \in \bar S \},
\end{equation}
where the limit of a sequence of past sets $\{P_n\}$ is given by
\begin{defn} 
\label{clim}
$Q = \lim \  P_n$ if  and  only if 
\begin{enumerate}[i)]
\item For each $x\in Q$, there is some $N$ such that $x \in P_n$ for
all $n>N$.
\item For each $x$ such that $I^-(x) \not \subset Q$, there is some $N$ such that $I^-(x) \not\subset P_n$
for all $n>N$.
\end{enumerate}
\end{defn}
The limit of a sequence of future sets is defined similarly.
There are, of course, many sequences $\{ P_n \}$ which do not converge
and for which no such limit exists. Since $M$ is a strongly causal spacetime, a lemma
below (lemma 2) will show that ({\it ii}) is equivalent to:

{\it ii) For each $x\notin Cl[Q]$, there is some $N$ such that $x \notin Cl[P_n]$
for all $n>N$.}

\noindent
However, we prefer the first definition above which is more directly associated with the causal 
structure.

These operations add only future
(past) boundary points to $\bar{S}$ (as the subscripts suggest), and
as an immediate consequence $Cl_{FB,PB}(\bar S) \cap M = \bar S \cap
M$. We then define sets
\begin{equation}
L^+(\bar{S}) = Cl_{FB}[\bar S \cup L^+_{IF}(\bar{S})],
\end{equation}
\begin{equation}
L^-(\bar{S}) = Cl_{PB}[\bar S \cup L^-_{IP}(\bar{S})].
\end{equation}
Note also that $L^\pm$ are not necessarily symmetric; $p \in
L^\pm(q)$ does not imply $q \in L^\mp(p)$. For example, in
figure~\ref{fig:badcaus}, $b \in L^-(a)$, but $a \notin L^+(b)$. This
implies $L^\pm$ are also not related to the causal future and past in
any causality on $\bar{M}$.
On the other hand, we have
$L^+(\bar{S}) \cap M = L^+_{IF}(\bar{S}) \cap M$,
$L^-(\bar{S}) \cap M = L^-_{IP}(\bar{S}) \cap M$, so these definitions
are just adding some boundary points to make $L^\pm_{IF,IP}$ look more
like closed sets.  

We can now define our topology:
\begin{defn}
\label{strongtop}
The topology $\bar{\cal T}$ on $\bar{M}$ is defined to be the coarsest
in which all the sets $\bar{M} \setminus L^\pm(\bar{S})$ are open for any
$\bar{S} \subset \bar{M}$.
\end{defn}
Note that the above is equivalent to stating that the collection of
$\bar M \setminus L^\pm(\bar S)$ form a sub-basis for the topology, as
the definition amounts to saying that all those sets and only those
sets which can be written as arbitrary unions and finite intersections
of the sets $\bar{M} \setminus L^\pm(\bar{S})$ are open. 

\subsection{$\bar{M}$ is $M$ with a boundary}
\label{emb}

Our first task is to understand the relation of this topology to the
manifold topology on $M$. We will show that the natural map $\Phi: M
\to \bar{M}$ is a homeomorphism from $M$ to $\Phi(M)$, and that the
image $\Phi(M)$ is dense in $\bar{M}$. This shows that with our
topology $\bar{\cal T}$, the completion $\bar{M}$ does equip $M$ with a boundary in a
topological sense.\footnote{Note however that $\bar{M}$ is not the
necessarily a compactification of $M$, as we do not prove that
$\bar{M}$ is compact in the topology $\bar{\cal T}$. For example, the
$\bar M$ for Minkowski space is non-compact due to the existence of 
spacelike infinity.} Thus, our topology satisfies the main requirement
we argued for in the introduction; it enables us to use $\bar M$ as a
tool for discussing limits in $M$. 

It is useful to introduce the notion of a future-closed set:
\begin{defn}
A set $S \subset M$ is future-closed in $M$ if and only if $\forall x
\in S$, $I^+(x) \subset S$.
\end{defn}
We define past-closed sets in $M$ and future- and past-closed sets in
$\bar{M}$ similarly. Note that a future-set is always future-closed.
Note that $L^+(\bar{S}) \cap M$ is future-closed in $M$ and $L^+(\bar S)$ is future-closed in $\bar M$.  We now prove
some useful lemmas (we omit dual results, where future and past are
exchanged).

\begin{lemma} \label{futclos}
For a future-closed set $F \subset M$, every point in the boundary of
$F$ (in $M$, using the topology of $M$) can be approached from the
future along some curve $\gamma \subset F$.
\end{lemma}
{\it Proof:} Since $M$ is strongly causal, every $x$ in the boundary
of $F$ has an open neighborhood $U$ which is both homeomorphic and
causally isomorphic to an Alexandrov neighborhood in Minkowski space.
This implies $I^+(x) \cap U \subset F$, and $x$ can be approached from
the future along $\gamma \subset I^+(x) \cap U \subset F$. $\Box$
\begin{lemma} \label{closed}
Given $\bar S \subset \bar M$, $L^+(\bar S) \cap M$ is a closed subset
of $M$.
\end{lemma}
{\it Proof:} Since $L^+(\bar S) \cap M = L^+_{IF}(\bar S) \cap M$ is a
future-closed set, any point $y$ in its boundary (in $M$) can be
approached from the future along a timelike curve $\gamma \subset
L^+(\bar S) \cap M$. As a result, $y$ has $I^+(y) = I^+ [\gamma]
\subset \cup_{\bar Q \in \bar S} Q^*$ and so $y \in L^+_{IF}(\bar
S)$. $\Box$

\begin{lemma} \label{IC}
The sets $I^\pm_C(\bar S)$ are open in $\bar M$ for every $\bar S
\subset \bar M$.
\end{lemma}
{\it Proof:} To show this, we will prove $I^+_C(\bar S) = \bar M
\setminus L^-(\bar S^+) $ for suitable $\bar S^+$. Let
\begin{equation}
\bar S^+ = \bar M \setminus I^+_{C}(\bar S).
\end{equation}
Then since $L^-(\bar S^+) \supset \bar S^+ = \bar M
\setminus I^+_{C}(\bar S),$ it is clear that $I^+_C(\bar S) \supset
\bar M \setminus L^-(\bar S^+)$.

Thus, we need only show $I^+_C(\bar S) \subset \bar M \setminus
L^-(\bar S^+)$, which is equivalent to $I^+_C(\bar S) \cap
L^-(\bar S^+) = \emptyset$. Now $I^+_C(\bar S) \cap L^-(\bar
S^+) = I^+_C(\bar S) \cap L^-_{IP}(\bar S^+)$, as no
past boundary point can belong to $I^+_C(\bar S)$, and $I^+_C(\bar S)
\cap \bar S^+ = \emptyset$ by definition. Furthermore, the
definition of $\bar S^+$ is equivalent to
\begin{equation}
\bar S^+ = \{ \bar P: P \cap R^* = \emptyset \ \forall \bar R
\in \bar S \}.
\end{equation}
Hence $(\cup_{\bar P \in \bar S^+} P) \cap R^* = \emptyset$ for
all $\bar R \in \bar S$. Any point $\bar Q \in L^-_{IP}(\bar S^+)$ has $Q \subset \cup_{\bar P \in \bar S^+} P$, so $Q \cap R^*
= \emptyset$ for all $\bar R \in \bar S$. Thus, $I^+_C(\bar S) \cap
L^-(\bar S^+) = I^+_C(\bar S) \cap L^-_{IP}(\bar S^+) = \emptyset$.  We conclude that,
\begin{equation}
\label{512C}
 I^+_C(\bar S) \subset [\bar M \setminus L^-(\bar S^+) ] . 
\end{equation}
$\Box$

We may now prove one of the crucial theorems in our work.
\begin{thm}
\label{homeo}
$M$ is homeomorphic to its image in $\bar M$.
\end{thm}
{\it Proof:} By lemma~\ref{closed} above (and the obvious dual) the
sets $L^\pm(\bar S) \cap M$ are closed sets in the topology of
$M$. Thus, the open sets of the induced topology are open sets in the manifold
topology on $M$. 

It remains to show that we include all the open sets in the manifold
topology. But from theorem \ref{chronmatch} we have the equality
$I^\pm(p) = I^\pm_C(p) \cap M$.  By lemma \ref{IC} these sets are open
in the topology induced on $M$ from $\bar M$.  Since $M$ is strongly
causal, these sets form a subbasis for the manifold topology and any
open set $U \subset M$ is of the form $\bar U \cap M$ for some $\bar
U$ open in $\bar M$. $\Box$

This theorem shows that the topology we are defining is compatible
with the pre-existing topology on $M$, in much the same way that we
saw that the chronology on $\bar{M}$ was compatible with the
pre-existing chronology on $M$ in section~\ref{chron}. In fact, we can
do slightly better:
\begin{thm} \label{opens}
Any subset $U \subset M$ is open if and only if $\Phi(U) \subset
\bar{M}$ is open in $\bar{M}$.
\end{thm}
{\it Proof:} Choose $x,y \in M$ such that $I^+(x) \cap I^-(y)$ lies in
a set $K \subset M$ which is compact in the manifold topology.
Consider $\bar P \in I^+_{C}(x) \cap I^-_{C}(y)$.  Then $\bar P$ is
attached by timelike curves through $M$ to $x$ and $y$.  Thus, either
$\bar P\in M$ or $\bar P$ is a TIP of some curve $\gamma$ (and also a
TIF of some $\gamma'$) in $I^+(x) \cap I^-(y)$.  But since $I^+(x)
\cap I^-(y)$ lies within a compact set in $M$, there are no such TIPs
or TIFs and $I^+_{C}(x) \cap I^-_{C}(y) = I^+(x) \cap I^-(y)$.  Such
sets are open in $\bar M$ by lemma \ref{IC}.  Finally, since any open
set in $M$ is the union of such Alexandrov sets, it is open in $\bar
M$.

On the other hand, if $\Phi(U)$ is open in $\bar M$, $U$ is open in
$M$ by theorem \ref{homeo}.  $\Box$

\subsubsection{Limits in $\bar M$}
\label{limit}

The main purpose in introducing a topology is that it allows us to
give some sequences in $M$ ideal points as endpoints. Thus, we can
discuss asymptotic behaviour in $M$ in terms of the points of $\bar
M$. We would now like to show that the limits defined by the topology
introduced above are consistent with the chronology. That is, we would
like to see that ideal points which are limit points in a casual sense
are also limit points in a topological sense. 

\begin{lemma}
\label{causlem}
For $x \in L^+(\bar{S})$ and any timelike curve $\gamma$ through $x$, 
we have $\bar Q \in L^+(\bar S)$ for any point of the form $\bar Q = (I^-[\gamma],Q^*) \in \bar M$. 
\end{lemma}
{\it Proof:} We have $Q^* \subset f(I^-[\gamma]) \subset I^+(x) \subset \cup_{\bar P
\in \bar S} P^*$, so $I^-[\gamma] \in L^+(\bar{S})$, unless $Q^* = \emptyset$.  In the latter case, consider a sequence of points $x_n \in
\gamma$ such that $\lim \ I^-(x_n) = I^-[\gamma]$. By the above
argument, $x_n \in L^+_{IP}(\bar{S})$, so $\bar{Q} \in Cl_{FB}[\bar{S}
\cup L^+_{IP}(\bar{S})] = L^+(\bar{S})$. $\Box$

\begin{thm} \label{endp}
For a timelike curve $\gamma$, any $\bar P \in \bar M$ of the form $\bar
P =(I^-[\gamma],P^*)$ is a future endpoint of $\gamma$ in $\bar M$.
\end{thm}
{\it Proof:} Since we have already shown that $M$ is homeomorphic to
its image in $\bar M$, the case where $I^-[\gamma]$ is a PIP is
trivial.  Thus we suppose it is a TIP. Parametrize the curve $\gamma$
by $\lambda$ increasing towards the future. We wish to prove that any
open set $U \subset \bar M$ (in our topology above) containing
$\bar P $ also contains a future segment of $\gamma$---that is, it
contains $x(\lambda)$ for $\lambda > \lambda_c$ for some
$\lambda_c$. It is sufficient to prove this for the subbasis that
generates the topology, so we may take $U = \bar M \setminus
L^\pm(\bar S)$.

So, suppose $\bar P \notin L^-(\bar S)$, which implies $\bar P
\notin L^-_{IP}(\bar S)$, since $P = I^-[\gamma]$ is an IP. Then
$I^-[\gamma]$ is not a subset of $\cup_{\bar Q \in \bar S} Q$.  But
$I^-[\gamma]$ is the union of $I^-(x)$ for $x \in \gamma$ and the
$I^-(x)$ form an increasing sequence.  Thus, there must be some
$\lambda_c$ such that $I^-[x(\lambda_c)]$ is not a subset of
$\cup_{\bar Q
\in \bar S} Q$.  As a result, $x(\lambda_c) \notin L^-(\bar
S)$. Furthermore, since $I^-[x(\lambda_1)] \subset I^-[x(\lambda_2)]$
for $\lambda_1 < \lambda_2$, $x(\lambda) \notin L^-(\bar S) \forall
\lambda > \lambda_c$.  Thus $\bar M \setminus L^-(\bar S)$ contains
$x(\lambda)$ for all $\lambda > \lambda_c$.

Now suppose $x \in L^+(\bar S)$ for some $x \in \gamma$. Then from
lemma \ref{causlem} $\bar P \in L^+(\bar S)$.  Thus, $\bar P \in
\bar M \setminus L^+(\bar S)$ implies $x \in \bar M \setminus L^+(\bar
S)$ for all $x \in \gamma$. $\Box$

This theorem shows that this topology satisfies a minimal criterion for
consistency with the causal structure.  An immediate corollary of this
result is that $M$ is a dense subset of $\bar{M}$, since every open
neighbourhood of a TIP $I^-[\gamma]$ contains points of $M$, namely
the future segment of $\gamma$. Thus, $\bar{M}$ is $M$ with a
boundary.

A more general relation between causal and topological limits can in
fact be proved, using the notion of limits introduced in definition
\ref{clim}:
\begin{thm} \label{plim}
If $P = \lim \  I^-(x_n)$ and $P^* = \lim \  I^+(x_n)$ for some $\bar P = (P,P^*) \in \bar M$, the
sequence $\{ x_n \} \subset M$ converges to $\bar{P}$ in the topology on
$\bar{M}$. 
\end{thm}
{\it Proof:} We need to show that the sequence $\{ x_n \}$ enters
every open neighbourhood of $\bar{P}$. As in the proof of
theorem~\ref{endp}, it is sufficient to consider the subbasis $\bar{M}
\setminus L^\pm(\bar{S})$. Equivalently, we wish to show that $x_n
\in L^\pm(\bar{S})$ for $n > N$ for some $N$ implies $\bar{P} \in
L^\pm(\bar{S})$. 

Now $x_n$ is an ordinary point, so $x_n \in L^-(\bar{S})$ implies
$I^-(x_n) \subset \cup_{\bar Q \in \bar S} Q$. If $P$ is non-empty, $P
=\lim \ I^-(x_n)$ implies that every $x \in P$ is in $I^-(x_n)$
for $n > N$; hence $x \in \cup_{\bar Q \in \bar S} Q$ for every $x \in
P$. That is, $P \subset \cup_{\bar Q \in \bar S} Q$, and hence $\bar{P} \in
L^-(\bar{S})$. If on the other hand $P$ is empty, $x_n \in
L^-_{IF}(\bar{S})$ and $P^* = \lim \ I^+(x_n)$ implies $\bar{P}$
is in $Cl_{PB}[L^-_{IF}(\bar{S})]$, and so $\bar{P} \in L^-(\bar{S})$.
The dual proof for futures is the same. $\Box$

\begin{figure}
\begin{center}
    \includegraphics[width=0.15\textwidth]{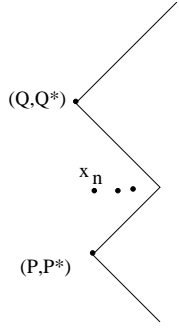}
\end{center}
\caption{An example where $P = \lim \ I^-(x_n)$ but we do not expect
the $\{ x_n \}$ converge to $\bar P$. Rather, the $\{ x_n \}$
intuitively converge to a point at `spacelike infinity' which is not
included in the causal completion $\bar M$. It is interesting to note
that although a point at spacelike infinity cannot have any spacetime
regions to its future or past, this does not imply that sequences $\{
x_n \}$ which go to spacelike infinity have $\lim \ I^\pm(x_n) =
\emptyset$. }  \label{fig:corner}
\end{figure}

In our approach, based on IP-IF pairs, these two theorems seem like an
appropriate statement of the consistency between topological and
causal limits. Like our definition of $\bar M$, theorem~\ref{plim}
uses the information in both pasts and futures.  There is an obvious
weaker version of the statement of theorem~\ref{plim}, which would
require that $\{ x_n \}$ converges to $\bar P$ if $P = \lim \
I^-(x_n)$ {\it or} $P^* = \lim \ I^+(x_n)$. There is no reason to
expect this to be generally satisfied. Indeed, figure~\ref{fig:corner}
provides an example where $P = \lim \ I^-(x_n)$ but we do not want the
topology to be such that the $\{ x_n \}$ converge to $\bar P$.  Indeed, they do not in our topology above, though they do converge to both $\bar P$ and $\bar Q$ in
the topology $\bar {\cal T}_{alt}$ introduced in appendix B. Returning to the topology $\bar{\cal T}$, since theorem~\ref{endp} (and
the obvious dual) makes reference only to pasts or futures, 
this weaker form will in fact be satisfied in some cases. For
example, in figure~\ref{fig:two}, any curve such that $A = I^-[\gamma]$ has
both $(A,B^*)$ and $(A,C^*)$ as limit points, even though $I^+(x_n)
\to B^* \cup C^*$ along $\gamma$. 

However, while the topology is thus consistent with the causality, it
may not give all the limits we would intuitively expect. For example,
in figure~\ref{fig:top}, the pasts of the $x_n$ will approach the past
of $q$, while the futures will approach the future of $p$. Hence, the
sequence $\{ x_n \}$ is not guaranteed to converge to either $p$ or
$q$. In fact, $q \notin L^+(\{ x_n \})$ and $p \notin L^-(\{ x_n \})$,
so $\bar M \setminus L^\pm(\{ x_n \})$ are open sets containing $q$
($p$) which the sequence never enters, and hence the sequence $\{ x_n
\}$ converges to neither of the points, whereas we would intuitively
expect it to converge to both. The same would be true if the $\{x_n \}$ were arrayed along a null line
from $p$ or $q$.  Now one could take the attitude that
our intuition is using additional information that is not contained in
the causal structure in this case; our argument above suggests that
the causal structure alone does not allow us to unequivocally say that
this sequence should or should not converge. However, there are some
somewhat arbitrary choices in our definition of the topology; in
appendix~\ref{app:weak}, we explore a slightly different alternative
definition of the topology which produces the more intuitively
satisfying answer in this example.

\subsection{Separation properties of $\bar{M}$}
\label{sep}

Having seen that the topology $\bar {\cal T}$ is compatible with the
manifold topology on $M$, we would like to investigate its separation
properties. We have previously argued that we should not assume that
the topology will be Hausdorff in general. Szabados has argued
in~\cite{szab1,szab2} that compatibility with the definition of the
boundary from the causal structure suggests that we impose a slightly
weaker separation condition, dubbed $T_c$, which asserts that every
causal curve has a unique endpoint in $\bar{M}$. However, in
figure~\ref{fig:two}, the points $(A,B^*)$ and $(A,C^*)$ are both
future endpoints of the same causal curve, so $T_c$ (and hence also
Hausdorffness) must fail\footnote{In general $T_c$ may also fail because, in analogy with figure \ref{fig:top}, not every null curve need have an
endpoint in $(\bar M, \bar {\cal T})$.}. Hence, in our approach where $\bar M$ is
constructed from IP-IF pairs, assuming $(\bar M,\bar{\cal T})$ is
Hausdorff or $T_c$ would be incompatible with demanding that the
topological notion of limits is consistent with the causal notion.

 \begin{figure}
\begin{center}
     \includegraphics[width=0.3\textwidth]{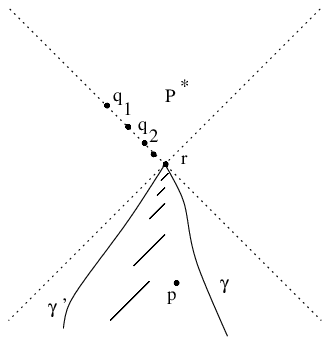}
\end{center}
\caption{An example in which $\bar M$ is not $T_1$.  Every closed
set containing $\bar P = (I^-[\gamma],P^*)$ also contains $\bar P' = (I^-[\gamma'],\emptyset)$. } \label{fig:Szab3}
\end{figure}

While our topology will not satisfy Szabados'
$T_c$ condition in its original form, his arguments suggest that we
should demand that the only future endpoints of a curve $\gamma$ in
$\bar M$ are the points of the form
$(I^-[\gamma],P_i^*)$. Unfortunately, this is again false in our
topology. In fact, even more worryingly, in general it does not even
satisfy the $T_1$ condition (which states that for any two points
$\bar P, \bar Q \in \bar M$, there are open sets ${\cal O}_P, {\cal
O}_Q \in \bar{\cal T}$ containing $\bar P$ but not $\bar Q$ and
vice-versa). Figure \ref{fig:Szab3} shows an example in which every
closed set containing $\bar P$ also contains $\bar P'$, and thus every open set
containing $\bar P'$ contains $\bar P$.  The example comes from figure 3 of
\cite{szab2}.  In this example, an infinite set of null lines
(including the limit point $r$) have been removed from Minkowski space
in such a way that future-directed timelike curves may cross from left
to right but not from right to left.  Thus, there are two interesting
TIPs, $P' = I^-[\gamma']$, $P = I^-[\gamma]$, with $P' \subset P$.  Note that there is a pair $\bar P =
(P,P^*)$ in $R_{pf}$, but not $(P', P^*)$, since $P' \subset P$ is not
maximal in any $p(P^*)$; rather, we have $\bar P' = (P', \emptyset) \in \bar M$.
One can easily see that $\bar P' \in L^-(\bar P)$.  In addition, $\bar P' \in
L^+(p)$ for any $p \in P$, since $I^-(q_n)$ converge to $P'$.  Thus, $\bar P'$
lies in any closed set containing $\bar P$, and $\bar P$ lies in any open set containing $\bar P'$.

This example shows that we cannot hope to prove that the causal
completion $\bar M$ with the topology introduced above satisfies even the
$T_1$ separation axiom in general. Our course in the remainder of this
section will first be to prove what we can, seeing how far we can get
with partial results which serve to isolate the problems with the
topology somewhat, and then discuss what we should do about them.

It follows immediately from theorem~\ref{opens} that any two points
$p,q \in M$ are $T_2$ separated in $\bar{M}$. We can prove that any
point $x \in M$ is $T_2$ separated from any ideal point by a series of
theorems:
\begin{thm}
\label{sepend}
Any point $p$ in a causal curve $\gamma$ is $T_2$ separated from
any point of the form $\bar P = (I^-[\gamma], P^*) \in \bar M$.
\end{thm}
{\it Proof:} To find an open set containing $p$ but not $I^-[\gamma]$,
use Theorem~\ref{opens} to choose any open $U \subset M$ containing
$p$.  For later convenience, take $U \subset I^-(q)$ for some $q \in
\gamma$ with $I^-(q) \supset I^-(p)$.

On the other hand, to find an open set containing $I^-[\gamma]$ but
not $p$, consider $\bar M \setminus L^-(q)$ for the same $q$ used
above. Clearly $p \in L^-(q)$, so $p \notin \bar M \setminus L^-(q)$.
Note that $q \notin I^-(q)$ since $M$ is causal.  Thus, $I^-(\gamma)
\not \subset I^-(q)$ and $\bar P \notin L^-_{IF}(q)$.    But since $I^-[\gamma]$
is a nonempty IP, $\bar P$ does not lie on the past boundary and we also
have $\bar P \in \bar M \setminus L^-(q)$.   

Since $I^-(q) \cap [\bar M \setminus L^-(q)] = \emptyset$ we have shown
that $p$ and $\bar P$ are $T_2$ separated.  $\Box$

\begin{thm}
If $K \subset M$ is compact in $M$, then $\Phi(K)$ is closed in $\bar M$.
\end{thm}
{\it Proof:} For each TIP $P$, choose some timelike curve $\gamma$
which generates $P$ (i.e., with $P = I^-[\gamma]$) and choose some $x
\in \gamma$ such that $I^+(x) \cap K = \emptyset$ (this is possible by
strong causality).  Recall that $I^+_C(x)$ is open in $\bar M$ by lemma \ref{IC}, and that
this set contains any $\bar P = (I^-[\gamma], P^*) \in \bar M$.
Taking the union of such sets
creates an open set which contains all TIPs but which does not
intersect $K$. We can similarly construct an open set which contains
all TIFs and does not intersect $K$. Finally, from theorem~\ref{opens}
and the Hausdorff property of $M$, given any $y \in M \setminus K$, we
can find $U_y \ni y$ open in $\bar M$ for which $U_y \cap K =
\emptyset$. The union of all such sets must be open, and is clearly
$\bar M \setminus \Phi(K)$.  Thus, $\Phi(K)$ is closed in $\bar M$. $\Box$

An immediate corollary of this theorem is that any point $p \in M$ is
$T_2$ separated from any ideal point $\bar P$: we may choose any $K
\ni x$ compact in $M$ and any open $U \subset K$ containing $x$. Then
$\bar M \setminus \Phi(K)$ is an open set containing $\bar P$ but not $x$,
while $U$ contains $x$ but not $\bar{P}$, and $U \cap \bar M \setminus
\Phi(K) = \emptyset$.

Thus, points $p \in M$ are $T_2$ separated from each other and from
ideal points. It remains to consider the separation of ideal
points among themselves. First, consider the ideal points which are
future endpoints of the same timelike curve $\gamma$ in the causal
sense. 
\begin{thm}
For any timelike curve $\gamma$, all the ideal points $\bar P_i$ with
$P_i = I^-[\gamma]$ are $T_1$ separated.
\end{thm}
{\it Proof:} We must have $P^*_i \subset f(I^-[\gamma])$ for
each $i$, and furthermore we cannot have $P^*_i \subset P^*_j$ for $i
\neq j$, or $P^*_i$ would not be maximal. Thus, $\bar P_i \notin
L^+(\bar P_j)$ for $i \neq j$. Hence $\bar M \setminus L^+(\bar P_j)$
is an open set containing $\bar P_i$ but not $\bar P_j$. $\Box$

Hence, the failure of $T_1$ is not associated with our decision to
ascribe several future endpoints to the same curve in some cases. 

Let us consider the more general case, not covered by this
proof. There are four different cases (omitting those which just swap
$\bar P$ and $\bar Q$ or pasts and futures): 1. Both $P$ and $P^*$
nonempty, both $Q$ and $Q^*$ nonempty; 2. Both $P$ and $P^*$ nonempty,
$Q^* = \emptyset$; 3. $P^* = \emptyset$, $Q^* = \emptyset$; 4. $P^* =
\emptyset$, $Q = \emptyset$.
\begin{thm}
Consider points $\bar P \neq \bar Q$ in $\bar M$.
\begin{enumerate}
\item If both $P$ and $P^*$ are nonempty and both $Q$ and $Q^*$ are
nonempty, $P \neq Q$, $P^* \neq Q^*$, the two points are $T_c$
(and thus $T_1$) separated. 
\item If both $P$ and $P^*$ are nonempty but $Q^* = \emptyset$, the
two points are $T_0$ separated.
\item If $P^* = \emptyset$ and $Q^* = \emptyset$, the two points are
$T_1$ separated. 
\item If $P^* = \emptyset$ and $Q = \emptyset$, the two points are
$T_1$ separated. 
\end{enumerate}
\end{thm}
{\it Proof:} \begin{enumerate}
\item For $T_c$ separation to fail, there must be some curve $\gamma$
such that $I^-[\gamma] = P$ and every open neighbourhood of $\bar Q$
contains a future segment of $\gamma$. This implies $\bar Q \notin
\bar M \setminus L^-(\bar{P})$ and $\bar Q \notin \bar M \setminus
L^+(x)$ for any point $x \in \gamma$, as $\gamma$ does not enter
either of these sets towards the future. That is, to violate $T_c$, we
must have $\bar Q \in L^-(\bar P)$ and $\bar Q \in L^+(x)$ for all $x
\in \gamma$, where $P = I^-[\gamma]$. Now $\bar Q \in L^-(\bar P)$
implies $Q \subset I^-[\gamma]$, and $\bar
Q \in L^+(x)$ implies $Q^* \subset I^+(x)$ for any $x \in
\gamma$. This implies $p(I^+(x)) \subset p(Q^*)$, and as $I^-[\gamma]
= \cup_{x \in \gamma} I^-(x) \subset \cup_{x \in \gamma}
p(I^+(x))$, we have $Q \subset P \subset p(Q^*)$, which contradicts
the assumption that $(Q,Q^*) \in R_{pf}$, since $P \neq Q$.  Hence the
points are $T_c$ (and thus $T_1$) separated.    
\item $L^+(\bar Q) = \bar Q$; hence $\bar M \setminus L^-(\bar Q)$
contains $\bar P$ but not $\bar Q$, and the two points are $T_0$
separated. Note that for $T_1$ to fail, we must have $\bar Q \in
L^-(\bar P)$, so $Q \subset P$.
\item $L^+(\bar P) = \bar P$; hence $\bar M \setminus L^+(\bar P)$
contains $\bar Q$ but not $\bar P$. Similarly, $\bar M \setminus
L^+(\bar Q)$ contains $\bar P$ but not $\bar Q$. Hence the two points
are $T_1$ separated.
\item $L^+(\bar P) = \bar P$; hence $\bar M \setminus L^+(\bar P)$
contains $\bar Q$ but not $\bar P$. Similarly, $\bar M \setminus
L^-(\bar Q)$ contains $\bar P$ but not $\bar Q$. Hence the two points
are $T_1$ separated.  $\Box$
\end{enumerate}

There are two interesting points to this theorem. First of all, it
shows that violations of $T_1$ happen only in cases which have the
flavor of figure~\ref{fig:Szab3}: there must be a pair of ideal points
$\bar P$ and $\bar Q$ such that $Q^* = \emptyset$ and $Q \subset P$
(or dually $Q = \emptyset$ and $Q^* \subset P^*$). Thus, the
conditions under which violations of $T_1$ can arise are restricted
and reasonably well-understood. Furthermore, we see that all the
`unexpected' violations of $T_c$ (where $\bar Q$ is a future endpoint
of a curve $\gamma$ for which $Q \neq I^-[\gamma]$) involve cases
where at least one of the pasts and futures involved is empty. Hence,
the points which fail to be appropriately separated would not have
been identified even if we had adopted the quotient approach,
completing $R_{pf}$ to the equivalence relation $\bar R_{pf}$.

Now we should consider what we want to do about the violations of
$T_1$. In the example of figure~\ref{top}, it was clear that
preserving the causal structure was more important than ensuring a
Hausdorff topology. The example of figure \ref{fig:Szab3} is less
clear, and one might consider it more natural to identify the two
ideal points that fail to be $T_1$ separated. 

Imposing such identifications in general will once again lead to
problems with the chronology.  One can construct an example where
problems arise by introducing barriers of the kind indicated by the solid lines in 
figure~\ref{fig:Szab3} in both the
past and the future of some point, in a somewhat more complicated version of the example in appendix
\ref{app:ex}, containing even more regions. Thus, any identifications will have to be performed on a
case by case basis. The question then is whether it is better to live
with the completion $\bar M$, which is at least generally defined,
even if some points are not $T_1$ separated, or to identify points as
much as we can without introducing new chronological
relations?\footnote{A third possibility would simply be to excise from
$\bar M$ any point of the form $(Q, \emptyset)$ or $(\emptyset, Q^*)$
which is not $T_1$ separated from some IP-IF pair $\bar P$. This would
avoid any problems with the chronology, but it would imply that any
curve $\gamma'$ such that $I^-[\gamma'] = Q$ would not have a future
endpoint in $\bar M$.  In particular, $\bar P$ is not an endpoint of
this curve: $I^-[\gamma']$ is a proper subset of  $P$, so $\bar P \notin L^-(\gamma')$,
and $\bar M \setminus L^-(\gamma')$ is an open set containing $\bar P$
which $\gamma'$ never enters.} If one does not want to impose any
identifications, one has a further choice of approach. Either one
regards $T_1$ separation as a desirable but not essential property,
and works with this completion $\bar M$ for any strongly causal
manifold, or one can insist that $T_1$ separation is important, and
only accept the completion for those cases where the topology
$\bar{\cal T}$ satisfies $T_1$. Note that the latter approach is
closer to that adopted in~\cite{szab2}, where the discussion was
restricted to manifolds satisfying certain `asymptotic causality
conditions' in addition to strong causality. We will leave these
issues for future discussion. In the unlikely event that such
problematic cases are encountered in practical applications, the
physics of the situation should provide better guidance than we can give here.

\section{Homogeneous Plane Waves}
\label{plane}

Plane wave spacetimes \cite{Brinkmann,BPR,JH,AP,EK,exact} provide a
very interesting set of examples to which these
techniques may be applied. The simplest non-trivial cases are smooth homogeneous
plane waves. Except in two special cases which are conformally flat,
the presence of a homogeneous Weyl tensor implies that these
spacetimes cannot be conformally embedded into a compact region of a
smooth manifold.\footnote{We thank Gary Horowitz for this argument:
The infinite conformal rescaling would require the Weyl tensor of the
spacetime in which we embed to diverge.} Thus, the only available
tool for discussing the asymptotic structure of these spacetimes is
the causal boundary approach. Application of this technique was first
considered in~\cite{beyond}, where it was argued that for homogeneous
plane waves satisfying any positive energy condition, the causal
boundary generically consists of a single null line.  This was in
agreement with the results found in \cite{bn} for a conformally flat
special case by conformally embedding the planewave in $S^n \times R$.
See \cite{toapp} for comments on how the generic behaviour differs
from the conformally flat case when the behaviour of spacelike curves
is considered. The causal boundary of plane waves has been further
studied in~\cite{hubeny:cb}.

We will now discuss the construction of the causal boundary for the
plane wave using the approach we have introduced, showing that we
recover the results of~\cite{beyond}. Homogeneous plane waves
illustrate the differences between the various ways of constructing
$\bar M$ in an interesting way, so we will also briefly compare the
results to those of the previous quotient constructions. The results
of~\cite{beyond} are also obtained when the Budic-Sachs-Szabados
identifications are used; using the original GKP recipe for
identifications, on the other hand, yields no identifications between
TIPs and TIFs.

\begin{figure}
\begin{center}
    \includegraphics[width=0.2\textwidth]{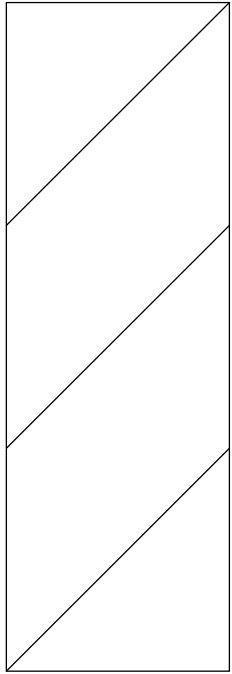}
\end{center}
\caption{A 1+1 dimensional slice of the
Einstein static universe along a great circle of the sphere.  Left
and right sides of the diagram are identified.  The winding null ray
indicates the conformal boundary of the conformally flat plane wave
under the Berenstein-Nastase conformal embedding \cite{bn}.}\label{ESU}
\end{figure}

The class of smooth homogeneous plane waves considered
in~\cite{beyond} is described by the metric 
\begin{equation} \label{pwave}
ds^2 = -2 dx^+ dx^- - (\mu_1^2 x_1^2 + \ldots + \mu_j^2 x_j^2 - m_1^2
y_1^2 - \ldots - m_{n-j}^2 y_{n-j}^2 ) dx_+^2 + dx^i dx^i + dy^a dy^a,
\end{equation} 
where $\mu_1 \geq \mu_2 \geq \ldots \geq \mu_j$. We assume there is at
least one $x^i$ (that is, $j \neq 0$), excluding a case with all
negative directions which violates the positive energy conditions and
has a different causal boundary. The special case $j=n$, $\mu_1 =
\mu_2 = \ldots = \mu_n$ is conformally flat, and it was shown
in~\cite{bn} that this conformally flat plane wave can be embedded
into the Einstein static universe, as depicted in
figure~\ref{ESU}. This conformal embedding into $S^n \times R$ maps
this plane wave onto the entire space, save for a single null geodesic
$N$.

In~\cite{beyond}, the definitions of section~\ref{rev} were applied to find
that the TIPs and TIFs for (\ref{pwave}) are of the form 
\begin{equation}
P_{x^+_0} = \{ x: x^+ < x^+_0    \}
\ \ \ {\rm and} \ \ \ 
P^*_{x^+_0} = \{ x: x^+ > x^+_0    \}.
\end{equation}
The common future and pasts for these TIPs and TIFs are given
by\footnote{In this step, a key point is that the dimension of the
spacetime is greater than two, so that the null line has codimension
greater than one.  Thus, to see (\ref{PWidents}) from figure \ref{ESU}
it is essential to recall that timelike curves may pass `around and
beyond' $N$.}
\begin{equation}
\label{PWidents}
f(P_{x^+_0}) = P^*_{x^+_0 + \pi/\mu_1} \ \ \ {\rm and} \ \ \
p(P^*_{x^+_0}) = P_{x^+_0 - \pi/\mu_1}.
\end{equation}
Here we allow $x^+_0$ to range over $[-\infty,+\infty]$, including the endpoints
$\pm \infty$.
We therefore see that the set of ideal points will consist of the
pairs $(P_{x^+_0}, P^*_{x^+_0 + \pi/\mu_1})$, as claimed
in~\cite{beyond}. This gives a single one-dimensional null line as the
conformal boundary of the spacetime. 

In the conformally flat case, this causal boundary agrees with the
boundary in the conformal compactification depicted in figure
\ref{ESU}. Furthermore, it is clear that the causal structure and
topology on $\bar{M}$ introduced in sections~\ref{struct}
and~\ref{top} agree with the natural causal structure and manifold
topology on the Einstein static universe.

Things are not so satisfactory in some of the alternative
approaches. If one considers the original scheme of GKP, we have
\begin{eqnarray}
\label{PWext}
P_{x^+} \in (P^*_{x^+_1})^{ext} &\ \mbox{ iff }& x^+ \le x^+_1, \cr 
 P^*_{x^+} \in (P_{x^+_1})^{ext} &\ \mbox{ iff }& x^+ \ge x^+_1.
\end{eqnarray}
All of these sets are open, and $(P^*_{x^+_1})^{ext} \cap
(P_{x^+_2})^{ext} = \emptyset$ for $x^+_1 \le x^+_2 $.  Thus
$P_{x_1^+}$ and $P^*_{x_2^+}$ are $T_2$ separated in the GKP topology
on $M^\sharp$ for $x_2^+ > x_1^+$.  Furthermore,
\begin{eqnarray}
\label{PWint}
P_{x^+} \in (P_{x^+_1}^*)^{int} &\ \mbox{ iff }& x^+ > x^+_1, \cr
P^*_{x^+} \in (P_{x^+_1})^{int} &\ \mbox{ iff }& x^+ < x^+_1
\end{eqnarray}
and $(P^*_{x^+_1})^{int} \cap (P_{x^+_2})^{int} = \emptyset$ for
$x^+_1 \ge x^+_2 $.  As a result, $P_{x^+_1}$ and $P^*_{x^+_2}$ are
$T_2$ separated for $x_1^+ \ge x_2^+$.  Thus, no TIP is identified
with any TIF.

Finally, since $(P_{x^+})^{int} \cap (P_{x^+})^{ext} = \emptyset$, any
two TIFs are $T_2$ separated.  Similarly, any two TIPs are $T_2$
separated and we see that no identifications are induced
whatsoever. Thus, the original GKP procedure produces a very different
answer than \cite{beyond}, which notably fails to agree with the
result of a natural conformal compactification in the conformally flat
case.

In the approach of \cite{szab1}, the identifications are determined by
$R_{pf}$, so this approach will similarly identify $P_{x^+_0}$ with
$P^*_{x^+_0 + \pi/\mu_1}$, and the point-set $\bar M_{id}$ agrees with
the result of conformal compactification. However, the topology on
$\bar M_{id}$ is less satisfactory. In~\cite{szab1}, a subset $U
\subset \bar M$ is open if and only if its inverse image in $M^\sharp$
is open.  Inspired by the example in \cite{diff3}, consider in
particular the set $U_{x^+_0} \in \bar M$ containing all ideal points
and all points in $M$ with $x^+ \neq x^+_0$.  Note that its inverse
image in $M^\sharp$ is the open set $(P_{x^+_0})^{ext} \cup
(P^*_{x^+_0})^{ext} \cup (P_{x^+_0})^{int} \cup (P^*_{x^+_0})^{int}$,
which again contains all ideal points and $I^\pm(p)$ for $p \in M$
with $x^+(p) \neq x^+_0$.  Because $U_{x^+_0}$ contains all ideal
points, no sequence $\{ p_n \}$ contained within the $x^+=x^+_0$ null
plane of $M$ can converge to an ideal point.  This is in sharp
contrast to the natural manifold topology on the conformal
compactification in the conformally flat case depicted in figure \ref{ESU}.

Lastly, since the plane wave is a causally continuous spacetime, we
can apply the approach of Budic and Sachs~\cite{budic}. This involves
the same identifications as~\cite{szab1} in this case, so it will
produce the same $\bar M_{id}$. The interesting difference now comes
in the topology. Since the chronology and causality relations defined
on $\bar M_{id}$ in~\cite{budic} will agree with the natural
chronology and causality on the ESU, the generalised Alexandrov
topology adopted there will agree with the manifold topology on the
conformal compactification.

Thus, the previous generally-applicable quotient approaches fail to
reproduce the structure of the conformal compactification for the
conformally flat case. One can compare the approaches in the cases of de Sitter and anti-de Sitter space
as well.  There all constructions produce the same point sets as the conformal approach.  However, much as
above only the topology of definition \ref{strongtop} and that of Budic and Sachs~\cite{budic}  agree with the usual conformal compactification.  

In fact, the result that $I^\pm_C(\bar P) \in \bar {\cal T}$ means that the Budic-Sachs topology will agree with 
$\bar {\cal T}$ in many interesting examples; for example, in figure \ref{fig:top}.
However, the Budic-Sachs construction can only be applied for causally continuous 
manifolds.
Thus, ours is the only generally applicable approach that achieves this result.

Although plane waves are not asymptotically simple, the fact that our new approach
achieves agreement with the conformal compactification in an
interesting example where previous approaches had failed to do so
seems to argue in its favor, and gives us more confidence in its
description of the causal boundary for the more general smooth
homogeneous plane waves.

\section{Discussion}
\label{disc}

We have developed the approach (initiated by~\cite{Geroch}) to constructing ideal points for a
spacetime from causal information.  This approach
has the advantage of being much more generally applicable than
alternative approaches to the construction of a boundary for
spacetime. In addition, it leads to a unique completion of any strongly causal spacetime.
This contrasts with the situation in a conformal approach, where uniqueness is not guaranteed
outside of the asymptotically simple context.  The causal
approach has previously been revised and
elaborated in various ways~\cite{budic,szab1,szab2,racz} in response to technical
difficulties encountered in specific
examples~\cite{diff1,diff2,diff3}.

We have argued that since this approach is based on the chronology
$\ll$ on $M$, which determines the IPs and IFs on which the
construction of the completion is based, an essential requirement in
such an approach is that there exists a chronology $\ll_C$ on $\bar M$
such that the restriction of $\ll_C$ to $M$ is isomorphic to $\ll$. We
observed that this requirement was not satisfied for arbitrary
strongly causal manifolds $M$ in any of the existing approaches based
on identifying equivalence classes of IPs and IFs. 

We introduced a new construction, where the points of $\bar M$ are
given by pairs $(P,P^*)$ of an IP with an IF. $\bar M$ includes every
pair such that $P$ and $P^*$ satisfy the relation $R_{pf}$. 
In general, the causal approach seems to construct a `big'
completion. Perhaps this should not be surprising, since it involves
relatively little information about the relations between the
points. 
We have
shown that we can define a chronology $\ll_C$ on this $\bar M$ which
satisfies the above requirement. Thus, this approach provides a
satisfactory causal completion for arbitrary strongly causal
spacetimes. We also investigated the definition of a causality on
$\bar M$; we can define a causality from the chronology so that $\bar
M$ is a causal space, but this causality has some unsatisfactory
features. 

In addition to the chronology, the structure with which we most wish to equip
$\bar M$ is a topology. This is difficult to do; the only natural
topology defined solely by the chronology is the Alexandrov topology,
but as argued in~\cite{Geroch}, this is not sufficiently strong to
provide suitable open neighbourhoods of all ideal points. The
definition of a different topology from the causal structure
necessarily involves some technical choices. 

We have introduced a topology $\bar{\cal T}$ on $\bar M$, based on
defining suitable closed sets from the causal structure. This topology
is stronger than the Alexandrov topology. We showed that $\bar M$ in
this topology corresponds to $M$ equipped with a boundary in a
suitable topological sense. Thus, we can use points in $\bar M$ to
describe limits of sequences in $M$. We have also shown that the
endpoints this topology defines are consistent with the causal
structure. Thus, this topology satisfies the key requirements we would
like to impose. However, it does not produce all the endpoints one
intuitively expects. This led us to introduce an alternative topology
$\bar{\cal T}_{alt}$ in appendix B, based on slightly different technical choices. $\bar{\cal T}_{alt}$
 leads to more limit points in interesting examples but sometimes goes a bit far.
In particular, as discussed in appendix B the sequence $\{x_n\}$ in figure \ref{fig:corner}
converges to both $\bar P$ and $\bar Q$ in $(\bar M,\bar  {\cal T}_{alt})$.  Thus we do not
regard either of the topologies we have introduced here as the final
answer; rather, they exhibit some of the desirable features achievable
in a particular approach to defining  causality-based
topologies. We hope this will act as an inspiration to further work. 

The main failing of our approach is that $(\bar M, \bar{\cal T})$ does
not generally satisfy separation axioms stronger than $T_0$ (this
problem becomes considerably worse with the topology $\bar{\cal
T}_{alt}$, which does not even satisfy $T_0$).  This seems to be just
the way it is; it is hard to see how modifications of the topology
could correct the problem illustrated in figure~\ref{fig:Szab3}. It
remains possible, however, that future work will introduce a superior
topology.  It would be surprising to us if these technical problems with
topological separation actually played a role in practical examples.

To illustrate the our approach, we briefly discussed the causal
completions for the homogeneous plane waves discussed
in~\cite{beyond}, filling in some of the details omitted in our
previous treatment. The subtleties of the differences between the
various approaches play an important role in this case. We found that
the previous generally-applicable approaches fail to agree with the
conformal compactification of~\cite{bn} in the conformally flat case.
The original procedure of~\cite{Geroch} would produce no
identification between TIPs and TIFs, producing a different point-set
structure for the completion, while the approach of~\cite{szab1} gives
the right point-set structure but produces a topology which does not
agree with the natural manifold topology in the conformal
compactification. Our new approach, and the approach for causally
continuous spaces of~\cite{budic}, successfully reproduce the
conformal compactification.  Similar results are obtained for both de Sitter and anti-de Sitter space.  The fact  that our new approach
is the only one for which i) $M \rightarrow \bar M$ is generally known to yield a dense embedding and ii) agreement with the conformal compactification  in such interesting examples is achieved 
seems to argue in its favor.

\acknowledgments 

DM would like to thank Joshua Goldberg, James B. Hartle, Gary Horowitz, and Rafael
Sorkin for useful discussions.  Part of this work was performed while
DM was a visitor at the Perimeter Institute (Waterloo, Ontario), and
part was performed while visiting Centro Estudios Cientificos
(Valdivia, Chile).  He would like to thank both institutes for their
hospitality.  DM was supported in part by NSF grant PHY00-98747 and by
funds from Syracuse University. SFR was supported by an EPSRC Advanced
Fellowship.

\appendix

\section{An example in 2+1 dimensions}
\label{app:ex}

\begin{figure}
\begin{center}
    \includegraphics[width=0.8\textwidth]{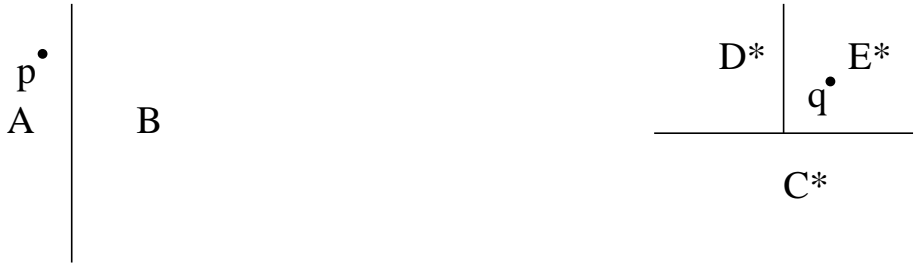}
\end{center}
\caption{Slices $t=-1/2$ (left) and $t=1/2$ (right), showing the cuts
made to produce our example. The cuts block the chronological relation
between the points $p$ and $q$, $p \not\ll q$. However, in the
construction of $\bar M_{id}$, all of the TIPs and TIFs shown are
identified to form a single ideal point $\bar P$. We then have $p
\ll_C \bar P$ and $\bar P \ll_C q$, but $p \not\ll_C q$, so $\ll_C$ is
not transitive.}\label{fig:ex}
\end{figure}

We have claimed that the main advantage of our new approach to the
construction of $\bar M$ is that it allows us to define a chronology
$\ll_C$ on $\bar M$ such that the natural chronology $\ll$ on $M$ is
isomorphic to $\ll_C$ on the image of $M$ in $\bar M$. It is therefore
important to exhibit an example where this fails to be true in the
previous approach based on quotients. 

We can easily construct such an example starting from a $2+1$
dimensional flat space. We delete from the space the following
surfaces: $\{x=0, -1 \leq t \leq 0\}$, $\{y=0, 0 \leq t \leq 1\}$, and $\{x=0,
y>0, 0 \leq t \leq 1\}$. This gives us a space with two TIPs $A,B$ and
three TIFs $C^*,D^*,E^*$ associated with the origin, illustrated in
figure~\ref{fig:ex}. From these, we can construct the pairs $(A,C^*)$,
$(A,D^*)$, $(B,C^*)$ and $(B,E^*)$ which are all in $R_{pf}$.

If we complete $R_{pf}$ to an equivalence relation $\bar{R}_{pf}$, we
will then have $A \sim B \sim C^* \sim D^* \sim E^*$; that is, all
five sets will be identified to form a single ideal point $\bar P$ in
$\bar M_{id}$. This might na\"\i vely seem desirable, since we started
with a single point at the origin before the surgery. However, it
leads to problems with the chronology. 

In $\bar M_{id}$, we can define a relation $\ll_C$ in essentially the
same way as we did on $\bar M$, and it still agrees with the
chronology on $M$. However, it will fail to be a chronology in our
example. The points $p \in A$ and $q \in E^*$ in figure~\ref{fig:ex}
are not chronologically related in $M$, and $p \not\ll_C q$, in
agreement with the chronology on $M$. However, $p \ll_C \bar P$ and
$\bar P \ll_C q$. Hence, $\ll_C$ fails to define a transitive relation
on $\bar M_{id}$. One might try to complete it to obtain a transitive
relation $\ll_{tr}$, but one will then have $p \ll_{tr} q$, and the
chronology will fail to agree with that on $M$. Furthermore, it is
easy to construct slightly more complex examples where this transitive
closure will lead to $p \ll_{tr} p$, so the relation $\ll_{tr}$ can
fail to be anti-reflexive.

Our interpretation is that although it may be na\"\i vely appealing to
identify all the TIPs and TIFs to have a single ideal point at the
origin, we cannot do so if we want $\bar M$ to have a compatible
chronology, because the identifications lose too much
information. Instead, we must define $\bar M$ so that the past or
future of any ideal point is a {\it single} IP or IF. In our
definition, we will have the pairs $(A,C^*)$, $(A,D^*)$, $(B,C^*)$ and
$(B,E^*)$ as distinct points of $\bar M$. We will have $p \ll_C
(A,C^*)$, $p \ll_C (A,D^*)$, and $(B,E^*) \ll_C q$, but $(A,C^*)
\not\ll_C (B,E^*)$ and $(A,D^*) \not\ll_C (B,E^*)$, so there is no
problem with transitivity.

\section{Alternative topology}
\label{app:weak}

Here we introduce a second topology ${\bar {\cal T}}_{alt}$ which is
similar to ${\bar {\cal T}}$, but which is defined using a different
closure operation.  This will serve to illustrate the importance of
the technicalities in our definition of the topology, showing that the
definition of a novel topology based purely on the causal structure is a
subtle and difficult problem. We can prove stronger results relating
limits to ideal points in this topology, and we will see that it has
more intuitively satisfying behaviour for limits in the example of
figure~\ref{fig:top}. On the other hand, adopting this topology
considerably worsens the problems with separation and leads to less
satisfactory results in the example of figure \ref{fig:corner}. 

The new closure operations $Cl_\pm$ are
\begin{equation}
Cl_{+}(\bar S) = \bar S \cup \{ \bar Q \in \bar M : Q \neq \emptyset,  
Q = \lim \  P_n \ {\rm for} \ \{\bar P_n\} \in \bar S \},
\end{equation}
\begin{equation}
Cl_{-}(\bar
S) = \bar S \cup \{ \bar Q \in \bar M : Q^* \neq \emptyset, Q^* = \lim \ 
P_n^* \ {\rm for} \ \{\bar P_n\} \in \bar S \},
\end{equation}
where the limit of a sequence of sets $\{P_n\}$ is again given by
definition \ref{clim}. We then define sets
\begin{equation}
L_{alt}^+(\bar{S}) = Cl_{+}[L^+_{IF}(\bar{S})],
\end{equation}
\begin{equation}
L_{alt}^-(\bar{S}) = Cl_{-}[L^-_{IP}(\bar{S})],
\end{equation}
\begin{equation}
L_{alt}(\bar{S}_1,\bar S_2) = Cl_{+}[L^-_{IP}(\bar{S_1}) \cap
L^+_{IF}(\bar S_2)] \cup Cl_{-}[L^-_{IP}(\bar{S_1}) \cap L^+_{IF}(\bar
S_2)].
\end{equation}
Note that as before $L_{alt}^\pm$ need not be symmetric; $p \in
L_{alt}^\pm(q)$ does not imply $q \in L_{alt}^\mp(p)$. For example,
in figure~\ref{fig:badcaus}, $b \in L_{alt}^-(a)$, but $a \notin
L_{alt}^+(b)$. Thus $L_{alt}^\pm$ are still not related to the causal
future and past in any causality on $\bar{M}$.

The essential difference between this and the previous definition of
$L^\pm$ is that while the closures $Cl_{FB,PB}$ only added points of
the form $(P,\emptyset)$, $(\emptyset,P^*)$ respectively, belonging to
the two exceptional classes in our construction of $\bar M$, the new
closure operations $Cl_\pm$ can also add points with non-empty pasts
{\it and} futures (in the general class in the construction of $\bar
M$) to $L_{alt}^\pm$. Thus, $L^\pm_{alt}(\bar S)$ can contain points
whose futures (pasts) are not subsets of the future (past) of $\bar
S$. That $Cl_\pm$ nonetheless provide reasonable closures of
$L^+_{IF}(\bar S), L^-_{IP}(\bar S)$ is in some part justified by
lemma \ref{newCl} below, which shows that $Cl_\pm$ add only ideal
points to $L^+_{IF}(\bar S), L^-_{IP}(\bar S)$.  Thus
$L_{alt}^+(\bar{S}) \cap M = L^+_{IF}(\bar{S}) \cap M$ and
$L_{alt}^-(\bar{S}) \cap M = L^-_{IP}(\bar{S}) \cap M$.

As a final comment, note that $Cl_+ [L^-_{IP}(\bar S)] = L^-_{IP}(\bar
S)$.  Together with the dual result, this shows
$L_{alt}(\bar{S}_1,\bar S_2) \subset L_{alt}^-(\bar S_1) \cap
L_{alt}^+(\bar S_2)$.  However, $L_{alt}(\bar{S}_1,\bar S_2)$ and
$L_{alt}^-(\bar S_1) \cap L_{alt}^+(\bar S_2)$ need not be equal.
For example, in figure \ref{top}, if we consider an ideal point $\bar
R$ on the vertical line connecting $p$ and $q$, both $p$ and $q$ will
lie in $L_{alt}^+(\bar R)$.  Similarly, both points lie in
$L_{alt}^-(\bar R)$.  However, choosing two such points $\bar R_1
\ll_C \bar R_2,$ neither $p$ nor $q$ lie in $L_{alt}(\bar R_1, \bar
R_2)$.  Note that the introduction of the closed sets $L_{alt}(\bar
R_1, \bar R_2)$ is essential to guarantee that $p,q$ do not lie in
every closed set containing any such $\bar R$, which helps to ensure
that this example has reasonable separation properties.  More will be
said about the separation properties of this topology in section
\ref{sep2} below.

We can now define our alternative topology.
\begin{defn}
The topology $\bar{{\cal T}}_{alt}$ on $\bar{M}$ is defined to be the
coarsest in which all the sets $\bar{M} \setminus
L_{alt}^\pm(\bar{S})$, $\bar{M} \setminus L_{alt}(\bar{S}_1, \bar
S_2)$ are open for any $\bar{S}, \bar S_1, \bar S_2 \subset \bar{M}$.
\end{defn}

This topology is very similar to $\bar{\cal T}$.  In fact, we will be
able to carry over many of the key results proven in section \ref{top}
directly to ${\bar {\cal T}}_{alt}$.  Important tools in this
process are the following two lemmas:

\begin{figure}
\begin{center}
\includegraphics[width=0.3\textwidth]{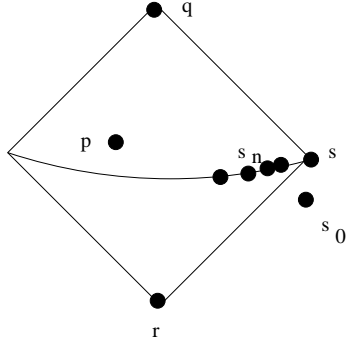}
\end{center}
\caption{ The point $s_0$ is in the past
of $s$, and thus in the past of all but a finite number of the $s_n$, but $I^-(s_0) \not \subset I^-(p)$.}
\label{fig:alex}  
\end{figure}
\begin{lemma} \label{newCl}
For $p \in M$, $p \in L_{alt}^+(\bar S)$ implies $p \in L^+_{IF}(\bar S)$.
\end{lemma}
{\it Proof:} Consider $p \in L_{alt}^+(\bar S) \cap M$.  Since $M$ is
strongly causal, we may choose some open neighborhood $U \ni p$ which
is causally isomorphic to Minkowski space.  Furthermore, choose any $q
\gg p, q\in U$ and any $r \ll p, r \in U$.

Now suppose that
$I^-(p) = \lim P_n$ for $\bar P_n \in L^+_{IF}(\bar S)$.  We will show
that all but a finite number of $P_n$ are of the form $I^-[p_n]$
for $q \gg p_n \gg r$.  Since this holds for any $q,r$ as above, we will have
shown $p_n \rightarrow p$ in the Alexandrov topology of $M$.  But this agrees with the manifold topology due to 
strong causality.  Since $p_n = \bar P_n 
\in L^+_{IF}(\bar S) \cap M$ and this set is closed in M by lemma \ref{closed}, we have $p \in L^+_{IF}(\bar S)$.

To complete the proof, we start with the observation of \cite{Geroch}
that $P_n = I^-[\gamma_n]$ for some timelike $\gamma_n$.  Now $r \in
I^-(p)$, so $r \in P_n$ for all but a finite number of $P_n.$ Thus, we
may take $\gamma_n \ni r$ for all these curves so that such $\gamma_n$
enter the Alexandrov neighborhood $I^+(r) \cap I^-(q)$.

But consider any $\gamma_n$ which exits this Alexandrov neighborhood.
Let $S(r,q)$ be the sphere on which the future null cone of $r$
intersects the past null cone of $q$.  There must be some $s_n \in
S(r,q) \cap I^-[\gamma_n]$.

Now suppose that more than a finite number of the $\gamma_n$ exit the
Alexandrov neighborhood.  Since $S(p,q)$ is compact, the set $\{s_n\}$
has some accumulation point $s$.  But this means that there is some
$s_0 \ll s$, $I^-(s_0) \not \subset I^-(p)$ for which $I^-(s_0) \in I^-[\gamma_n]$ for
more than a finite number of $\gamma_n$.  This would contradict the
condition that $I^-(p) = \lim I^-[\gamma_n]$, so we conclude that only
a finite number of $\gamma_n$ can exit $I^+(r) \cap I^-(q)$.  Since
this neighborhood is contained in $U$ which is causally isomorphic to
Minkowski space, a timelike curve $\gamma_n$ which does not exit
$I^+(r) \cap I^-(q)$ must have an endpoint $p_n \in I^+(r) \cap
I^-(q)$.  Repeating the argument for arbitrary $q,r$ with $q \gg p \gg r$
one finds $p_n \rightarrow p$ and $p \in L^+_{IF}(\bar S)$ as
explained above. $\Box$

\begin{lemma} \label{newCl2}
For $p \in M$, $p \in L_{alt}(\bar S_1,\bar S_2)$ implies $p \in
L^-_{IP}(\bar S_1) \cap L^+_{IF}(\bar S_2)$.
\end{lemma}
{\it Proof:} This follows immediately from lemma \ref{newCl} and the
observation that $L_{alt}(\bar S_1,\bar S_2) \subset L^+_{alt}(\bar
S_1) \cap L^-_{alt}(\bar S_2)$. $\Box$

We may now quickly prove

\begin{thm}
\label{homeo2}
$M$ is homeomorphic to its image in $(\bar M, {\bar {\cal T}}_{alt})$.
\end{thm}
{\it Proof:} By lemmas~\ref{newCl} and \ref{newCl2} above, we see that
the sets $L^+_{IF}(\bar S) \cap M$ and $L^-_{IP}(\bar S) \cap M$
form a subbasis for the topology induced on $M$ by the embedding
in ($\bar M, {\bar \cal T}_{alt}$).
But these sets also form a subbasis for the topology induced on $M$
by the embedding in $(\bar M, {\bar {\cal T}})$ and in theorem
\ref{homeo} this latter induced topology was shown to agree with the manifold
topology on $M$. $\Box$

As with ${\bar {\cal T}}$, we can in fact
also prove:
\begin{thm} \label{weak_opens}
Any subset $U \subset M$ is open if and only if $\Phi(U) \subset
\bar{M}$ is open in $(\bar{M}, {\bar {\cal T}}_{alt})$.
\end{thm}
{\it Proof:} Let $\bar S^+_x = \bar M \setminus I^+_C(x)$ and note that $ \bar M \setminus I^+_{C}(x) =
\bar{S}^+_x \subset L_{alt}^-(\bar S^+_x)$ implies $\bar M \setminus
L_{alt}^-(\bar S^+_x) \subset I^+_{C}(x)$. Thus, $[\bar M \setminus
L_{alt}^-(\bar S^+_x)] \cap [\bar M \setminus L_{alt}^+(\bar S^-_y)]
\subset I^+_{C} (x) \cap I^-_{C} (y)$. On the other hand, from lemma
\ref{newCl} we have $I^+(x) \cap L_{alt}^-(\bar S_x^+) = \emptyset$,
so $\bar M \setminus L_{alt}^-(\bar S_x^+) \supset I^+(x)$ and $[\bar
M \setminus L_{alt}^-(\bar S^+_x)] \cap [\bar M \setminus
L_{alt}^+(\bar S^-_y)] \supset I^+(x) \cap I^-(y)$.

Consider $\bar P \in I^+_{C}(x) \cap I^-_{C}(y)$.  Then $\bar P$ is
attached by timelike curves through $M$ to $x$ and $y$.  Thus, either
$\bar P\in M$ or $\bar P$ is a TIP of some curve $\gamma$ (and also a
TIF of some $\gamma'$) in $I^+(x) \cap I^-(y)$.  In the case where
$I^+(x) \cap I^-(y)$ lies within a compact set in $M$, there are no
such TIPs or TIFs and $I^+_{C}(x) \cap I^-_{C}(y) = I^+(x) \cap
I^-(y)$.  As a result, when $I^+(x) \cap I^-(y)$ lies within a compact
set in $M$, $\bar M \setminus L_{alt}^-(\bar S^+_x) \cap \bar M \setminus
L_{alt}^+(\bar S^-_y) = I^+(x) \cap I^-(y)$. Since any open set in $M$ is
the union of such Alexandrov sets, it is open in $\bar M$. 

On the other hand, if $\Phi(U)$ is open in $\bar M$, $U$ is open in
$M$ by theorem \ref{homeo}.  $\Box$

Next, we would like to show that the TIPs and TIFs, which are limit
points in a casual sense, are also limit points in a topological
sense. Below we will consider causal (as opposed to merely timelike)
curves.  Thus, the results below for ${\bar {\cal T}}_{alt}$ are
stronger than their analogues for ${\bar {\cal T}}$. Recall that
$I^-[\gamma]$ for causal $\gamma$ was shown to be an IP in
\cite{Geroch}.
\begin{lemma}
\label{weak_causlem}
For $x \in L_{alt}^+(\bar{S})$ and any causal curve $\gamma$ through $x$, 
we have $\bar Q \in L_{alt}^+(\bar S)$ for any $\bar Q$ of the form $(I^-[\gamma], Q^*) \in M$. 
\end{lemma}
{\it Proof:} First suppose that $I^-[\gamma]$ is a PIP and that
$\gamma$ has endpoint $p \in M$.  By lemma \ref{newCl}, $x \in
L^+_{IF}(\bar S)$.  Thus $I^+(p) \subset I^+(x) \subset \cup P^*$ for
$\bar P \in L^+_{IF}(\bar S)$ and $p \in L^+_{IF}(\bar S)$.

Now consider the general case.  Since $I^-[\gamma] \neq \emptyset$, we
need only show that $I^-[\gamma] = \lim P_n$ for $\bar P_n \in
L^+_{IF}(\bar S)$.  To do so, consider a sequence of points $x_n \in
\gamma$ such that $\lim \ I^-(x_n) = I^-[\gamma]$. By the above
argument, $x_n \in L^+_{IF}(\bar{S})$, so $\bar Q \in
Cl_{+}[L^+_{IF}(\bar{S})] = L_{alt}^+(\bar{S})$. $\Box$

\begin{thm} \label{weak_endp}
For a causal curve $\gamma$, any $\bar P \in \bar M$ of the form $\bar
P =(I^-[\gamma],P^*)$ is a future endpoint of $\gamma$ in $\bar M$.
\end{thm}
{\it Proof:} Since we have already shown that $M$ is homeomorphic to
its image in $\bar M$, the case where $I^-[\gamma]$ is a PIP is
trivial.  Thus we suppose it is a TIP. Parametrize the curve $\gamma$
by $\lambda$ increasing towards the future. We wish to prove that any
open set $U \subset \bar M$ (in the topology $\bar {\cal T}_{alt}$)
containing $\bar P $ also contains a future segment of $\gamma$---that
is, it contains $x(\lambda)$ for $\lambda > \lambda_c$ for some
$\lambda_c$. It is sufficient to prove this for the subbasis that
generates the topology, so we may take $\bar U = \bar M \setminus
L_{alt}^\pm(\bar S)$ or $\bar U = \bar M \setminus L_{alt}(\bar S_1,
\bar S_2)$.

So, suppose $\bar P \notin L_{alt}^-(\bar S)$, which implies $\bar P
\notin L^-_{IP}(\bar S)$.  Then since $I^-[\gamma] \neq \emptyset$, it
is not a subset of $\cup_{\bar Q \in \bar S} Q$.  But $I^-[\gamma]$ is
the union of $I^-(x)$ for $x \in \gamma$ and the $I^-(x)$ form an
increasing sequence.  Thus, there must be some $\lambda_c$ such that
$I^-[x(\lambda_c)]$ is not a subset of $\cup_{\bar Q \in \bar S} Q$.
As a result, $x(\lambda_c) \notin L_{alt}^-(\bar S)$. Furthermore,
since $I^-[x(\lambda_1)] \subset I^-[x(\lambda_2)]$ for $\lambda_1 <
\lambda_2$, $x(\lambda) \notin L_{alt}^-(\bar S) \forall \lambda >
\lambda_c$.  Thus $\bar M \setminus L_{alt}^-(\bar S)$ contains
$x(\lambda)$ for all $\lambda > \lambda_c$.

Now suppose $x \in L_{alt}^+(\bar S)$ for some $x \in \gamma$. Then
from lemma \ref{weak_causlem} we have $\bar P \in L_{alt}^+(\bar S)$.
Thus, $I^-[\gamma] \in \bar M \setminus L_{alt}^+(\bar S)$ implies $x
\in \bar M \setminus L_{alt}^+(\bar S)$ for all $x \in \gamma$.

Finally, suppose that any final segment of $\gamma$ contains some $x$
in $L_{alt}(\bar S_1,\bar S_2)$.  Together with lemma \ref{newCl2}, the arguments above then show that
some final segment $\gamma'$ satisfies
$\gamma' \subset L^+_{IF}(\bar S_1) \cap L^-_{IP}(\bar S_2)$.  Thus we
need only choose $x_n \in \gamma'$ with $\lim I^-(x_n) =
I^-[\gamma']$ to show $\bar P \in L_{alt}(\bar S_1,\bar S_2)$.  $\Box$

An immediate corollary of this result is that $M$ is a dense subset of
$\bar{M}$, since every open neighbourhood of a TIP $I^-[\gamma]$
contains points of $M$, namely the future segment of $\gamma$. Thus,
$\bar{M}$ is again $M$ with a boundary in the topology $\bar{\cal T}_{alt}$. 

For $\bar{\cal T}_{alt}$, we can prove a stronger version of
theorem \ref{plim}: 
\begin{thm}
If $P \neq \emptyset$, $P = \lim \  I^-(x_n)$ or $P^* \neq \emptyset$,
$P^* = \lim \ I^+(x_n)$, the sequence $\{ x_n \} \subset M$ converges to $\bar{P}$
in the topology $\bar{\cal T}_{alt}$ on $\bar{M}$. 
\end{thm}
{\it Proof:} We need to show that the sequence $\{ x_n \}$ enters every
open neighbourhood of $\bar{P}$. As in the proof of
theorem~\ref{weak_endp}, it is sufficient to consider the subbasis
$\bar{M} \setminus L_{alt}^\pm(\bar{S})$, $\bar M \setminus
L_{alt}(\bar S_1, \bar S_2)$. Equivalently, we wish to show that when
there exists an $N$ such that $x_n \in L_{alt}^\pm(\bar{S})$ for all $n
> N$, we have $\bar{P} \in L_{alt}^\pm(\bar{S})$, and similarly for $L_{alt}(\bar S_1,\bar S_2)$.

Now $x_n \in M$ is an ordinary point, so $x_n \in L_{alt}^-(\bar{S})$
implies $I^-(x_n) \subset \cup_{\bar Q \in \bar S} Q$ by lemma
\ref{newCl}. Thus, if $P = \lim \ I^-(x_n)$, then every $x \in P$ is in
$I^-(x_n)$ for sufficiently large $n$; hence $x \in \cup_{\bar Q \in
\bar S} Q$ for every $x \in P$. That is, $P \subset \cup_{\bar Q \in
\bar S} Q$, and hence $\bar{P} \in L^-_{IP}(\bar{S}) \subset I^-(\bar
S)$ and $\bar P \in L^-_{alt}(\bar S)$. If $P^* = \lim \ I^+(x_n)$, then $x_n \in L_{alt}^-(\bar{S})$ 
implies $\bar{P} \in Cl_{-}[L^-_{IP}(\bar{S})]$.  Again, $\bar{P} \in
L_{alt}^-(\bar{S})$.  The dual proof for futures is the same.

Finally, suppose $\{ x_n \} \subset L_{alt}(\bar S_1,\bar S_2) =
L^+_{IF}(\bar S_1) \cap L^-_{IP}(\bar S_2).$ Then since either $P = \lim
\ I^-(x_n)$ or $P^* = \lim \ I^+(x_n)$, $\bar P$ must lie in one of the
two closures $Cl_\pm$ of $L^+_{IF}(\bar S_1) \cap L^-_{IP}(\bar S_2).$
Thus $\bar P \in L_{alt}(\bar S_1,\bar S_2)$.  $\Box$

This theorem means that $\bar{\cal T}_{alt}$ will require the addition of far fewer points at `spacelike infinity' to $\bar M$ in order to compactify the spacetime.

Thus, the general properties of limits in this topology are slightly better
than in the previous topology. Furthermore, in the example in
figure~\ref{fig:top}, with the new definitions employed above, both
$p$ and $q$ are in $L^\pm(\{ x_n \})$, so there will be no open
neighbourhoods of $p$ or $q$ which the $x_n$ never enter --- the sequence $\{ x_n \}$
has both $p$ and $q$ as limit points, as we would intuitively
expect. Unfortunately, however, this topology goes a bit too far; in
figure~\ref{fig:corner}, the sequence $\{ x_n \}$ converges to both
$\bar P$ and $\bar Q$. 

\subsection{Separation properties of $(\bar{M}, {\bar {\cal T}}_{alt})$}
\label{sep2}

We have obtained slightly better behaviour for limits in this
alternative topology. It is then perhaps not surprising to find that
we obtain weaker results concerning the separation properties. In
fact, there is no general result concerning separation at all: we can
immediately see that we will have to give up even on $T_0$ whenever we
have more than one point of $\bar M$ with the same past or
future. Consider, for example, the left spacetime shown in figure
\ref{fig:two}.  Because they share the same IP, the closed sets
containing $(A,B^*)$ are exactly the closed sets containing $(A,C^*)$.
Thus any open set containing one of these points also contains the
other. These points are not separated at all in this topology. Thus,
unlike in the topology $\bar{\cal T}$, with the topology $\bar{\cal
T}_{alt}$, our decision to introduce distinct points of $\bar M$ for
each IP-IF pair is responsible for problems with the separation.

Some better control is obtained for points in $M$.  For example, it
follows immediately from theorem~\ref{weak_opens} that any two points
$p,q \in M$ are $T_2$ separated in $\bar{M}$.  Furthermore, we can
prove that any point $x \in M$ is $T_1$ separated from any ideal point
through the following lemma.
\begin{lemma}
\label{point}
Any point $p \in M$ defines a closed set $\{ p \} = L_{alt}(p,p)
\subset \bar M$.
\end{lemma}
{\it Proof:} Consider any $\bar Q = (Q,Q^*) \in L^+_{IF}(p) \cap
L^-_{IP}(p)$.  We have $Q^* \subset I^+(p)$ and $Q \subset I^-(p)$.
Thus $Q \subset I^-(p) \subset p(I^+(p)) \subset p(Q^*)$ and, since
$Q$ is maximal in $p(Q^*)$ we must have $Q = I^-(p)$.  Similarly, $Q^*
= I^+(p)$ and $p = \bar Q$.  But since $I^\pm(p)$ are related only to each other by $R_{pf}$,
 the closure operations $Cl_\pm$ can add no points:
$L_{alt}(p,p) = L^+_{IF}(p) \cap L^-_{IP}(p)$. $\Box$

It follows that
\begin{thm}
\label{sepideal}
Any point $p \in M$ is $T_1$ separated from any ideal point $\bar Q$.
\end{thm}
{\it Proof:} Lemma \ref{point} tells us that $\bar M \setminus \{p\}$ is
an open set containing $\bar Q$ and not $p$, while
theorem~\ref{weak_opens} tells us that any open $ U \subset M$
containing $p$ is an open set in $\bar M$ containing $p$ but not $\bar
Q$. $\Box$

Similarly, any points $\bar P = (P,P^*)$, $\bar Q = (Q,Q^*)$ with $P
\neq Q$, $P^* \neq Q^*$ are $T_1$ separated.  Here $P,P^*,Q,Q^*$ must be non-trivial.

\begin{figure}
\begin{center}
\includegraphics[width=0.6\textwidth]{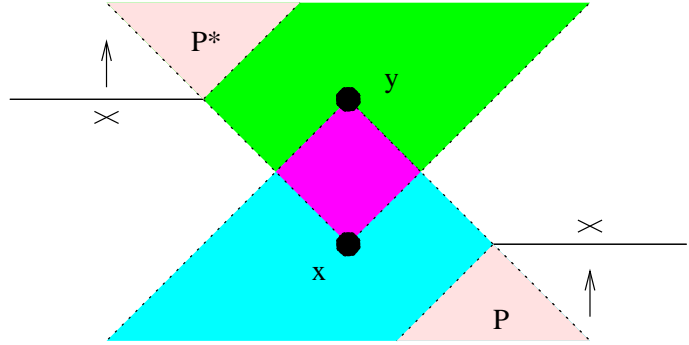}
\end{center}
\caption{The arrows indicate that the lower side of the right cut is
identified with the upper side of the left cut.  These
identifications include a right-left reflection, so that $(P,P^*) \in
R_{pf}$.  The crosses indicate that neither the upper side of the
right cut nor the lower side of the left cut is identified with
anything.  Since $P^* \subset I^+(x)$ and $P \subset I^-(y)$, $(P,P^*)
\in L_{alt}(x,y)$. An infinite set of such cuts and identifications can
lead to poor separation properties.}  \label{fig:sick}
\end{figure}
 
However, in general interior points and ideal points need not be $T_2$
separated.  The example in figure \ref{fig:sick} illustrates the
general idea.  In this example $(P,P^*)$ lies in $L_{alt}(x,y)$.  
This by itself causes no great harm -- one may simply work with
$z_1,z_2 \in I^+(x) \cap I^-(y)$ so that $L_{alt}(z_1,z_2)$ is just the
set $\{ p \in M : z_1 \ll p \ll z_2 \}$.  But in similar examples with an
infinite number of cuts and identifications, there may be a point $z$
such that some pair resembling $(P,P^*)$ in figure \ref{fig:sick} lies
in $L_{alt}(x,y)$, $L_{alt}^+(x)$, and $L_{alt}^-(y)$ for all $x \ll z, y
\gg z$.

In this case it would follow that $z$ is not $T_2$ separated from
$(P,P^*)$.  The argument is as follows: We have seen that any open set
containing $z$ contains some open $U \subset M$ with $U \ni z$.  An
open set that does not intersect $U$ must be constructed from $\bar M
\setminus L_{alt}(x,y)$, $\bar M \setminus L_{alt}^+(x)$, and $\bar M
\setminus L_{alt}^-(y)$.  But none of these open sets contain
$(P,P^*)$.  Thus, $z$ and $(P,P^*)$ are not $T_2$ separated.  With
some careful arrangement, one can find similar examples
in which a point on a timelike curve $\gamma$ is not $T_2$ separated
from an endpoint of this curve.

Clearly, this topology has undesirable separation properties. In
particular, its failure to recognize the distinction between different
IP-IF pairs which have the same future or past is
distressing. Furthermore, this is not the only problem: we have argued
that in sufficiently complicated examples, interior and boundary
points will fail to be $T_2$ separated in this topology (and of course
none of the problems with separation properties in $\bar{\cal T}$
encountered in section~\ref{sep} are corrected in this topology). 
Finally, it leads in figure \ref{fig:corner} to the convergence of $\{ x_n \}$ to both $\bar P$
and $\bar Q$.

Our main purpose in discussing this topology was to show that the
non-intuitive behaviour of certain limits in the topology $\bar{\cal
T}$ was not purely a consequence of working with the causal structure;
certain arbitrary choices in the definition of the topology also
affect these limits. The fact that $\bar {\cal T}_{alt}$ generally produces more limit points than
$\bar {\cal T}$ is a positive feature in some examples (such as figure \ref{fig:top}), but a negative feature in others (such as figure
\ref{fig:corner}).
One might hope that there is some more inspired
choice of definition, somehow intermediate between the ones we have
made, which would allow us to combine the good features of both
topologies.

\end{document}